\documentclass[pre,twocolumn,showpacs]{revtex4}
\usepackage{graphicx,amsmath}
\begin{document}

\title{Supercooled Liquids Under Shear: Theory and Simulation}

\author{Kunimasa Miyazaki}
\affiliation{Department of Chemistry and Chemical Biology,
Harvard University, 12 Oxford Street, Cambridge, MA 02138, U.S.A}

\author{David R. Reichman}
\affiliation{Department of Chemistry and Chemical Biology,
Harvard University, 12 Oxford Street, Cambridge, MA 02138, U.S.A}

\author{Ryoichi Yamamoto}
\affiliation{Department of Physics, Kyoto University, Kyoto 606-8502, 
           Japan,\\
PRESTO, Japan Science and Technology Agency,
              4-1-8 Honcho Kawaguchi, Saitama, Japan.} 

\begin{abstract} 
We analyze the 
behavior of 
supercooled fluids under shear 
both
theoretically and numerically. 
Theoretically, 
we generalize the mode-coupling theory of supercooled fluids to 
systems under stationary shear flow. 
Our starting point is the 
set of 
generalized fluctuating hydrodynamic equations
with a convection term. 
A nonlinear 
integro-differential equation 
for the intermediate scattering function is constructed.  
This theory is applied to a two-dimensional colloidal suspension.
The shear rate dependence of the intermediate scattering function and 
the shear viscosity is analyzed. 
We have also 
performed
extensive numerical simulations 
of
a two-dimensional binary liquid with 
soft-core interactions
near, but above, the glass transition
 temperature.  
Both theoretical and numerical results show:
(i) A drastic reduction of the structural relaxation time 
and the shear viscosity due to shear.  
Both the structural relaxation time and the viscosity
decrease as $\dot{\gamma}^{-\nu}$ with an exponent $\nu \leq 1$, 
where $\dot{\gamma}$ is the shear rate. 
(ii) 
Almost 
isotropic dynamics 
regardless of the strength of the anisotropic shear flow.

\end{abstract}
\pacs{64.70.Pf,61.43.Fs,05.70.Ln,47.50.+d}
\maketitle

\section{Introduction}
\renewcommand{\theequation}{\Roman{section}.\arabic{equation}}  
\setcounter{equation}{0}

Complex fluids such as 
colloidal
suspensions, polymer solutions
and granular 
fluids exhibit very diverse rheological 
behavior\cite{larson1999,doi1986}.  
Shear thinning is among the most 
well-known phenomena. 
Such behavior is predicted for simple liquids as
well\cite{naitoh1976,kirkpatrick1985},  
but the effect is too small to observe at temperatures well above the
glass transition temperature.  
For supercooled liquids, however, the situation is
different.  
Recently, 
strong shear thinning 
behavior
was observed by experiments
performed on 
soda-lime
silica glasses above the glass transition
temperature\cite{simmons1982}.  
Yamamoto {\it et al.} 
have done extensive computer simulations 
of a binary liquid with a
soft-core interaction\cite{yamamoto1998c}
near their glass-transition temperature 
and found non-Newtonian behavior. 
The same behavior was found for other systems 
such as Lennard-Jones mixtures\cite{barrat2000,angelani2002} 
and polymer
melts\cite{yamamoto2002}, too. 
For all cases, the
structural relaxation time 
and the shear viscosity decrease
as $\dot{\gamma}^{-\nu}$, where $\dot{\gamma}$ is the shear rate and
$\nu$ is an exponent which is less than but close to 1.
For such systems driven far from equilibrium, the 
parameter $\dot{\gamma}$
 is not a small perturbation
but plays a role 
similar to 
an intensive parameter 
which
characterizes the ``thermodynamic state'' of the system\cite{liu1998b}.  
Such rheological behavior is
interesting in its own right, but understanding the dynamics of
supercooled liquids in a nonequilibrium state is more important because
it has possibilities 
to shed light on an another typical and perhaps more 
important nonequilibrium problem, namely that of non-stationary aging. 
Aging is characterized by slow relaxation after a sudden
quench of temperature below the glass transition temperature.
In this case, the waiting time plays a similar role to 
(the inverse of) the shear rate. 
Aging behavior has been extensively studied in spin 
glasses (see Ref.\cite{cugliandolo2003} and references therein).
Aging is also observed in 
structural glasses\cite{kob2000}. 
There 
have been recent
attempts to study the aging of structural glasses 
theoretically\cite{latz2000} 
but 
no analysis and comparison to experiments or 
simulation results\cite{kob2000} 
have been presented 
due to the complicated nonstationary nature of the
problem.  

The relationship between 
aging and a 
driven, steady-state system
was considered using 
a schematic model based on the exactly solvable p-spin spin glass 
by Berthier, Barrat and Kurchan\cite{berthier2000b}. 
This theory naturally gives rise to effective temperatures.
The validity of their idea was tested numerically for supercooled 
liquids\cite{barrat2000}. 
There have also been 
attempts to observe aging 
by exerting shear on the system
instead of quenching the temperature\cite{viasnoff2002}.
Recently, there have been attempts to 
develop the mode-coupling theory for the sheared
glasses\cite{miyazaki2002,fuchs2002}. 
Fuchs and Cates
have developed the
mode-coupling theory for the sheared colloidal suspensions 
using projection operator techniques\cite{fuchs2002}. 
They have analyzed
a closed equation for the correlation function 
for a schematic model where the shear is exerted on all direction
equally (``The Isotropically Sheared Hard Sphere Model''). 
In their model, 
shear is turned on at 
the initial time and therefore
the dynamics 
are 
genuinely nonstationary. 

In this paper, we investigate the 
dynamics of supercooled fluids under shear
both theoretically and numerically for a realistic system.
We extend the standard mode-coupling theory (MCT) for supercooled fluids
and compare 
the solutions
with the numerical simulation results.
We mainly focus on the 
microscopic origin of the 
rheological behavior. 
The relationship
with the more generic
aging problem will not be discussed here. 
Since our goal is the investigation of shear thinning behavior,
we have 
neglected 
violations
of the 
fluctuation-dissipation theorem. 
We start with generalized fluctuating hydrodynamic equations with 
a convection term. 
Using several approximations,
we obtain a closed nonlinear equation for the 
intermediate scattering function for the sheared system.  
The theory is applicable to  both normal liquids and 
colloidal suspensions in the absence of hydrodynamic
interactions.  
Similar approach has been applied to the self-diffusion of the hard
sphere colloidal suspension at relatively low densities by Indrani {\it et
al.}\cite{indrani1995}. 
Extensive computer simulations are implemented for a two-dimensional
binary liquid interacting with 
soft-core 
interactions. 
The 
effects of shear on microscopic 
structure, 
the structural relaxation 
time, and rheological behavior are discussed. 
Results both from the theory and the 
simulation show good qualitative
agreement despite of the 
differences
 between the systems considered. 
Special attention 
is
paid 
to 
the directional
dependence of the 
structural relaxation. 

The paper is organized as follows: 
In the next section, we develop the MCT for 
sheared suspensions. 
Complexities which do not exist in 
mean-field spin glass models 
and how 
those complexities
should be treated are elucidated here. 
A possible way to explore the situation without 
invoking
the fluctuation-dissipation
theorem is also 
discussed. 
In Section III, the model and our simulation method are explained. 
The results both from theory and simulation are 
discussion
in Section IV. 
In Section V we conclude.

\section{Mode-Coupling Theory}
\label{sec:mct}
\renewcommand{\theequation}{\Roman{section}.\arabic{equation}}  
\setcounter{equation}{0}

We shall consider a two-dimensional colloidal suspension under 
a stationary 
shear flow given by 
\begin{equation}
{\bf v}_{0}({\bf r}) = \mbox{\boldmath$\Gamma$}\cdot{\bf r} 
= (\dot{\gamma} y, 0), 
\end{equation}
where $\dot{\gamma}$ is the shear rate and 
$(\mbox{\boldmath$\Gamma$})_{\alpha\beta}
=\dot{\gamma}\delta_{\alpha x}\delta_{\beta y}$
is the velocity gradient matrix.
Generalization to higher dimensions is trivial. 
We start with the generalized fluctuating hydrodynamic
equations\cite{das1985,kirkpatrick1986}. 
This is a natural generalization of the fluctuating hydrodynamics
developed by Landau and Lifshitz\cite{landau1959} to short wavelengths
where the intermolecular correlations becomes important. 
Fluctuating fields  for the number
density $\rho({\bf r}, t)$ and the
velocity ${\bf v}({\bf r}, t)$ of the colloidal 
suspension 
obey the following set of nonlinear Langevin equations
\begin{equation}
\begin{aligned}
&
\frac{\partial{\rho}}{\partial t}
= - \nabla\cdot(\rho{\bf v}),
\\
&
m\frac{\partial (\rho{\bf v})}{\partial t}
+m\nabla\cdot(\rho{\bf v}{\bf v}) 
= 
-\rho \nabla\frac{\delta {\cal F}}{\delta \rho}
-\zeta_{0}\rho({\bf v}-{\bf v}_{0}) + {\bf f}_{R},
\end{aligned}
\label{eq:hydrodynamcs}
\end{equation}
where $m$ is the mass of a single colloidal particle, 
${\cal F}$ is the free energy of the system and
$\zeta_{0}$ is a bare 
collective friction coefficient for colloidal
particles.
$\zeta_{0}$ has 
a
weak density and 
distance dependence
due to 
hydrodynamic interactions\cite{beenakker1984a} but we shall 
neglect these effects.
The friction  term is specific for the colloidal case. 
In the case of liquids, it should be replaced by 
a
stress term 
which is proportional to the gradient of the 
velocity field multiplied by the shear viscosity.
Both cases, however, lead to the same dynamical behavior 
on 
the 
long time
scales which are of 
interest here.
${\bf f}_{R}({\bf r}, t)$ is the random force which satisfies
the fluctuation-dissipation theorem of the second kind (2nd
FDT)\cite{footnote1};
\begin{equation}
\langle
{f}_{R,i}({\bf r}, t){f}_{R,j}({\bf r'}, t')
\rangle_{0}
= 2k_{\mbox{\tiny B}}T\rho({\bf r}, t)\zeta_{0}\delta_{ij}
\delta({\bf r}-{\bf r}')\delta(t-t')
\end{equation}
for $t\geq t'$, 
where $\langle\cdots\rangle_{0}$ is an average over the conditional
probability for a fixed value of $\rho({\bf r},t)$ at $t=t'$. 
Note that the random force depends on the density and thus
the noise is multiplicative. 
This fact makes 
a
mode-coupling analysis more involved as we discuss later
in this section. 
We 
assume
 that the 2nd FDT holds even in nonequilibrium
state since the correlation of the random forces are short-ranged and
short-lived, and thus the effect of the shear is expected to 
be negligible. 
The first term 
on 
the right hand side of the equation for the
momentum is the osmotic pressure term.
Here we assume that the free energy ${\cal F}$ 
is well
approximated by that of the equilibrium form and is given by 
the
well-known expression
\begin{equation}
\begin{aligned}
\beta{\cal F} 
\simeq 
&
\int\!\!\mbox{d}{\bf r}~ \rho({\bf r})\left\{ \ln\rho({\bf r})/\rho_0 -1 \right\}
\\
&
-\frac{1}{2}\int\!\!\mbox{d}{\bf r}_1\int\!\!\mbox{d}{\bf r}_2~
\delta\rho({\bf r}_1)c(r_{12})\delta\rho({\bf r}_2),
\end{aligned}
\end{equation}
where $\beta=1/k_{\mbox{\tiny B}} T$, $\rho_{0}$ is the average density, and 
$c(r)$ is the direct correlation function.
We have neglected 
correlations 
of 
more than three points, such as
the triplet correlation function $c_{3}({\bf r}_1,{\bf r}_2,{\bf r}_3)$, whose 
effect becomes important for the fluids with stronger
directional interactions
such as silica\cite{sciortino2001}. 
Under shear, it is expected that $c(r)$ will be distorted and should
be replaced by a nonequilibrium steady state form
$c_{\mbox{\tiny NE}}({\bf r})$, which is an anisotropic function of
${\bf r}$.  
It is, however, natural to expect that this distortion is 
small 
on
the molecular length 
scales
which 
play
 the most important role in the
slowing down of 
structural relaxation near the glass transition. 
The distortion of the structure under shear is given up to 
linear order in the shear rate by\cite{ronis1984}. 
\begin{equation}
S_{\mbox{\tiny NE}}({\bf k})
= 
S(k)
\left\{ 
1 + \frac{\hat{\bf k}\cdot{\mbox{\boldmath$\Gamma$}}\cdot\hat{\bf k}}{2kD_{0}}\frac{{\mbox{d}} S(k)}{{\mbox{d}} k}
\right\}
,
\label{eq:mct.Sk}
\end{equation}
where $S(k)$ and $S_{\mbox{\tiny NE}}({\bf k})$ is the static structure
factor in the absence and in the presence of the shear, respectively. 
$D_{0} = k_{\mbox{\scriptsize B}} T/\zeta_{0}$ is the diffusion coefficient in the dilute
limit, $k \equiv |{\bf k}|$, and $\hat{\bf k}\equiv {\bf k}/|{\bf k}|$. 
The direct correlation function in the Fourier representation of $c(r)$
is related to the structure factor by 
$nc(k)= 1 - 1/S(k)$. 
From eq.(\ref{eq:mct.Sk}), we find that the distortion due to the shear 
is characterized by the P\'{e}clet number defined by 
Pe$=\dot{\gamma}\sigma^2/D_0$, where $\sigma$ is the diameter of the
particle. 
Hereafter we shall neglect the distortion and use the direct
correlation function at equilibrium, assuming the P\'{e}clet number is
very small. 

We linearize eq.(\ref{eq:hydrodynamcs}) around the stationary state as
$\rho = \rho_{0}+\delta \rho$ and ${\bf v} = {\bf v}_{0} + \delta{\bf v}$,
where $\rho_{0}$ is the average density. 
Transforming to 
wave vector space,
we obtain the following equations;
\begin{equation}
\begin{aligned}
&
\left(
\frac
{\partial~}{\partial t}
- {\bf k}\cdot\mbox{\boldmath$\Gamma$}\cdot\frac{\partial ~}{\partial {\bf k}}
\right)
\delta\rho_{{\bf k}}(t) = \frac{ik}{m}J_{{\bf k}}(t),
\\
&
\left(
\frac{\partial ~}{\partial t}
- {\bf k}\cdot\mbox{\boldmath$\Gamma$}\cdot
 \frac{\partial ~}{\partial {\bf k}}
+ \hat{\bf k}\cdot\mbox{\boldmath$\Gamma$}\cdot\hat{\bf k}
\right)
J_{{\bf k}}(t)
\\
&
= 
-\frac{ik}{\beta S(k)}\delta\rho_{{\bf k}}(t)
-\frac{1}{m\beta}
\int_{{\bf q}}
i\hat{\bf k}\cdot{\bf q} c(q)\delta\rho_{{\bf k}-{\bf q}}(t)\delta\rho_{{\bf q}}(t)
\\
&
\hspace*{.5cm}
-\frac{\zeta_{0}}{m}J_{{\bf k}}(t) + {f}_{R{\bf k}}(t),
\end{aligned}
\label{eq:nonlinear}
\end{equation}
where $J_{{\bf k}}(t)=m\rho_{0}\hat{\bf k}\cdot\delta{\bf v}_{{\bf k}}(t)$ 
is the longitudinal momentum fluctuation, 
and 
$\int_{{\bf q}}\equiv \int\mbox{d}{\bf q}/(2\pi)^2$.
We have neglected the quadratic terms proportional to
${\bf J}_{{\bf q}}{\bf J}_{{\bf k}-{\bf q}}$. 
Note that eq.(\ref{eq:nonlinear}) does not contain
coupling to transverse momentum fluctuations even in the presence of
shear.

The direct numerical integration of eq.(\ref{eq:nonlinear}) is 
in principle possible but it is 
expensive and 
not theoretically enlightening\cite{lust1993}. 
Rather we shall construct the approximated closure for the
correlation functions, a so-called mode-coupling 
approximation\cite{gotze1992,zwanzig2001,kawasaki1970}. 
There are several approaches to derive 
mode-coupling equations, 
including the 
use 
of
the Mori-Zwanzig projection operator and 
a 
decoupling
approximation\cite{gotze1992}, 
or  
implementation of a 
loop expansion developed in the
context of the equilibrium critical phenomena and 
generalized 
to the dynamic case\cite{martin1973,kawasaki1970}.
Both approaches lead to 
essentially
the same equations if the system is at
equilibrium. 
Under nonequilibrium conditions, 
however, 
the loop expansion approach is more straightforward 
and flexible. 
We shall adopt the loop expansion approach to the nonlinear Langevin
equation with both multiplicative noise and the 
full
convection term.  
Eq.(\ref{eq:nonlinear}) can be cast to 
a general form of the nonlinear Langevin equation written as
\begin{equation}
\frac{\mbox{d} x_{i}}{\mbox{d} t } 
= \mu_{ij}x_j + \frac{1}{2}{\cal V}_{ijk}x_{j}x_{k} + f_{R,i}.
\label{eq:mct.nll}
\end{equation}
where 
${\bf x}(t)= (\delta\rho_{{\bf k}}(t), J_{{\bf k}}(t))$ is a field variable, 
$\mu_{ij}$ is the linear coefficient matrix, and the nonlinear coupling
coefficient 
${\cal V}_{ijk}$ is the vertex tensor which satisfies the symmetric relation 
${\cal V}_{ijk}={\cal V}_{ikj}$. 
Finally, $f_{R,i}(t)$ is the random force field which 
satisfies the 2nd FDT, 
\begin{equation}
\langle f_{R,i}(t)f_{R,j}(t') \rangle_{0}
= k_{\mbox{\scriptsize B}} L_{ij}({\bf x})\delta(t-t')
,
\end{equation}
where $L_{ij}({\bf x})$ is the ${\bf x}$-dependent Onsager coefficient which is
to be expanded to the lowest order as 
\begin{equation}
L_{ij}({\bf x}) = L_{ij}^{(0)} + L_{ij,k}^{(1)}x_k
\end{equation}
where $L_{ij,k}^{(1)} \equiv \partial L_{ij}({\bf x})/\partial x_k|_{\bf x=0}$.
Comparing eq.(\ref{eq:mct.nll}) with eq.(\ref{eq:nonlinear}), 
the elements of the linear coefficient matrix 
are given by 
\begin{equation}
\left\{
\begin{aligned}
&
\mu_{\rho_{{\bf k}}\rho_{{\bf k}'}}
= {\bf k}\cdot\mbox{\boldmath$\Gamma$}\cdot\frac{\partial ~}{\partial {\bf k}}
\delta_{{\bf k},{\bf k}'}
\\
&
\mu_{\rho_{{\bf k}}J_{{\bf k}'}}
= \frac{ik}{m}\delta_{{\bf k},{\bf k}'}
\\
&
\mu_{J_{{\bf k}}\rho_{{\bf k}'}}
= 
\frac{ik}{\beta S(k)}\delta_{{\bf k},{\bf k}'}
\\
&
\mu_{J_{{\bf k}}J_{{\bf k}'}}
= 
\left(
{\bf k}\cdot\mbox{\boldmath$\Gamma$}\cdot\frac{\partial ~}{\partial {\bf k}}
-
\hat{\bf k}\cdot\mbox{\boldmath$\Gamma$}\cdot\hat{\bf k}
\right)\delta_{{\bf k},{\bf k}'}
- \frac{\zeta_{0}}{m}\delta_{{\bf k},{\bf k}'}.
\end{aligned}
\right.
\label{eq:mct.muij}
\end{equation}
Nonzero elements of the vertex tensor are given by
\begin{equation}
\begin{aligned}
&
{\cal V}_{J_{{\bf k}}\rho_{{\bf k}'}\rho_{{\bf k}''}}
= -\frac{1}{\beta V}
\left\{  i\hat{\bf k}\cdot{\bf k}' c(k')+i\hat{\bf k}\cdot{\bf k}''c(k'') \right\}
\delta_{{\bf k},{\bf k}'+{\bf k}''}
,
\end{aligned}
\label{eq:mct.Vijk}
\end{equation}
where $V$ is the volume of the 
system. 
Finally,
\begin{equation}
\begin{aligned}
&
L^{(0)}_{J_{{\bf k}}J_{{\bf k}'}}
= \zeta_{0}\rho_{0} T V\delta_{{\bf k},-{\bf k}'}
\\
&
L^{(1)}_{J_{{\bf k}}J_{{\bf k}'},\rho_{{\bf k}''}}
= T\zeta_{0}\delta_{{\bf k}+{\bf k}',{\bf k}''}
.
\end{aligned}
\label{eq:mct.L}
\end{equation}
All other components are zero. 
In the above expressions, 
\[
 \delta_{{\bf k},{\bf k}'} \equiv \frac{(2\pi)^{2}}{V}\delta({\bf k}-{\bf k}')
\]
is the Dirac
delta function.
We construct the closure equation for the correlation function 
$C_{ij}(t, t')= \langle x_{i}(t)x_{j}(t')\rangle$. 
Since we are treating the stationary state, 
the time translation invariance holds and, thus, 
$C_{ij}(t, t')= C_{ij}(t-t')$. 
A general 
scheme for the loop expansion method for the Langevin equation
with multiplicative noise 
has been discussed by Phythian\cite{phythian1977}. 
Up to one-loop, the equation for the correlation function 
is written
as\cite{miyazaki2003b} 
\begin{equation}
\begin{aligned}
&
\frac{\mbox{d} C_{ij}(t-t')}{\mbox{d} t }
- \mu_{i\alpha}C_{\alpha j}(t-t')
-2k_{\mbox{\scriptsize B}} L_{i\alpha}^{(0)}{\hat{G}}^{\dagger}_{\alpha j}(t-t')
\\
&
= 
\int_{-\infty}^{t}\!\!{\mbox{d}} t_1~
\Sigma_{i\alpha}(t-t_1)C_{\alpha j}(t_1-t') 
\\
&
+ 
\int_{-\infty}^{t'}\!\!{\mbox{d}} t_1~
D_{i\alpha}(t-t_1){\hat{G}}^{\dagger}_{\alpha j}(t_1-t') 
\\
&
\frac{\mbox{d} {\hat{G}}_{ij}(t-t')}{\mbox{d} t } 
-\mu_{i\alpha}{\hat{G}}_{\alpha j}(t-t')
= 
\delta_{ij}\delta(t-t') 
\\
&
+
\int_{t'}^{t}\!\!{\mbox{d}} t_1~
\Sigma_{i\alpha}(t-t_1){\hat{G}}_{\alpha j} (t_1-t')
\end{aligned}
\label{eq:mct.msr1}
\end{equation}
with the memory kernels defined by 
\begin{equation}
\begin{aligned}
\Sigma_{ij}(t)
= 
&
{\cal V}_{i\alpha\beta}
{\hat{G}}_{\alpha\lambda}(t)C_{\beta\mu}(t)
{\cal V}_{\lambda\mu j}
\\
&
+
k_{\mbox{\scriptsize B}} 
{\cal V}_{i\alpha\beta}
{\hat{G}}_{\alpha\lambda}(t){\hat{G}}_{\beta\mu}(t)
L^{(1)}_{\lambda\mu, j}
\\
D_{ij}(t)
= 
&
\frac{1}{2}
{\cal V}_{i\alpha\beta}
C_{\alpha\lambda}(t)C_{\beta\mu}(t)
{\cal V}_{j \lambda\mu}
\\
&
+
2k_{\mbox{\scriptsize B}}{\cal V}_{i\alpha\beta}{\hat{G}}_{\alpha\lambda}(t)C_{\beta\mu}(t)
L^{(1)}_{j \lambda, \mu}
\\
&
+
2k_{\mbox{\scriptsize B}} L^{(1)}_{i\alpha,\beta}{\hat{G}}^{\dagger}_{\alpha\lambda}(t)C_{\beta\mu}(t)
{\cal V}_{j \lambda, \mu}
.
\end{aligned}
\label{eq:mct.Sigma-D}
\end{equation}
In these expression, we have introduced the propagator defined by 
\begin{equation}
{\hat{G}}_{ij}(t-t') 
= \left\langle \frac{\delta x_{i}(t)}{\delta f_{R,j}(t')}\right\rangle
.
\end{equation}
${\hat{G}}_{ij}^{\dagger}(t-t') \equiv {\hat{G}}_{ji}(t'-t)$ is the conjugate of
the propagator. 
Eq.(\ref{eq:mct.msr1}) together with eq.(\ref{eq:mct.Sigma-D}) 
are 
so-called mode-coupling 
equations.

Without further assumption, one needs to solve the coupled equation
for the correlation function and the propagator. 
The propagator can be eliminated, however, if the fluctuation-dissipation
theorem of the first kind (1st FDT) is valid. 
The 1st FDT relates the correlation function to the response function,
$\chi_{ij}(t)$ via the equation
\begin{equation}
\chi_{ij}(t)
=-\frac{\theta(t)}{k_{\mbox{\scriptsize B}} T}
\frac{\mbox{d} C_{ij}(t)}{\mbox{d} t } 
,
\label{eq:msr.1stFDT}
\end{equation}
where $\theta(t)$ is the heaviside function. 
Note that the response function $\chi_{ij}(t)$ is generally neither the
same as, nor proportional to, the propagator ${\hat{G}}_{ij}(t)$. 
$\chi_{ij}(t)$ represents the response of the system to 
the perturbation via external fields or through the boundary, whereas
${\hat{G}}_{ij}(t)$ represents the response to 
a 
random force field. 
If the Langevin equation is linear, both are the same, but 
this is 
not true
in general for the nonlinear 
Langevin 
equation. 
To one-loop,
 to be consistent with the mode-coupling
approximation, the response function is written as\cite{ma1975,miyazaki2003b} 
as 
\begin{equation}
\begin{aligned}
&
\chi_{ij}(t-t')
=
\frac{1}{T}{\hat{G}}_{ik}(t-t')\{ {\mbox{\boldmath$M$}}^{(0)} + {\mbox{\boldmath$L$}}^{(0)} \}_{kj}
\\
&
+
\frac{1}{T}
\int\!\!{\mbox{d}} t_{1}~
{\hat{G}}_{ik}(t-t_1){\cal V}_{k\alpha\beta}
{\hat{G}}_{\alpha\lambda}(t_1-t')C_{\beta\mu}(t_1-t')
\\
&
\times
\{ {\mbox{\boldmath$M$}}^{(1)} + {\mbox{\boldmath$L$}}^{(1)} \}_{\lambda j,\mu}
,
\end{aligned}
\label{eq:mct.chi}
\end{equation}
where the tensors $M^{(0)}_{ij}$ and $M^{(1)}_{ij,k}$ are defined by 
$M_{ij}({\bf x}) \simeq M^{(0)}_{ij} + M^{(1)}_{ij,k}x_{k}$. 
$M_{ij}({\bf x})$ 
is the matrix which is defined by 
the reversible part of the nonlinear Langevin equation. 
In general, we can express 
the 
Langevin equation as
\begin{equation}
\frac{\mbox{d} x_{i}}{\mbox{d} t } 
=  v_{i}({\bf x})
 + M_{ij}({\bf x})\frac{\delta S}{\delta x_{j}} 
 + L_{ij}({\bf x})\frac{\delta S}{\delta x_{j}} + f_{R,i},
\end{equation}
where $S$ is the entropy of the whole system and 
$v_{i}({\bf x})$ is a term which 
originates
 from the nonequilibrium
constraint such as the convection for the sheared system. 
From 
this 
definition, the reversible term does not contribute to the
entropy production and thus the matrix $M_{ij}({\bf x})$ is antisymmetric. 
For the system considered here, the non-zero elements are given by 
\begin{equation}
M_{\rho_{{\bf k}}J_{{\bf k}'}}
= -M_{J_{{\bf k}'}\rho_{{\bf k}}}
= -ik T
  \left( \rho_{0}V\delta_{{\bf k},-{\bf k}'}+\delta\rho_{{\bf k}+{\bf k}'}  \right)
\end{equation}

If the system is 
in 
equilibrium, one may eliminate the propagator in
favour of the correlation function in eq.(\ref{eq:mct.msr1}) by using 
the 1st FDT, eq.(\ref{eq:msr.1stFDT})
\cite{deker1975,bouchaud1996,miyazaki2003b}. 
If the system is 
out of
equilibrium, one has to solve the coupled
equations (\ref{eq:mct.msr1}). 
For the case of the 
$p$-spin-glass model 
with an external drive, 
Berthier {\it et al.} have 
analyzed the equation similar to
(\ref{eq:mct.msr1})\cite{berthier2000b}. 
They have found that there is a systematic deviation from the 1st FDT
which 
reminiscent of violations of FDT that occur during aging.
Extensive 
computer simulation supports these
results\cite{barrat2000}. 
A
similar analysis for 
real fluids 
is 
necessary but it is
inevitably more involved due to
the complicated tensorial nature of the non-linear coupling. 
In this paper, we shall not focus on the fundamental problem 
of 
the validity of the fluctuation-dissipation theorem
and assume that the 1st FDT is valid.
We shall show that even with this simplification, the theory can explain
qualitative several aspects of 
the dynamical behavior 
very well. 
Using the 1st FDT, as in equilibrium case, one can eliminate
${\hat{G}}_{ij}(t)$ 
from eq.(\ref{eq:mct.msr1}) and the final expression is given 
by\cite{miyazaki2003b}
\begin{equation}
\begin{aligned}
&
\frac{\mbox{d} C_{ij}(t-t')}{\mbox{d} t } 
- \mu_{ik}C_{kj}(t-t') 
-2k_{\mbox{\scriptsize B}} L^{(0)}_{ik}{\hat{G}}^{\dagger}_{kj}(t-t')
\\
&
= 
\int_{t'}^{t}{\mbox{d}} t_{1}~M_{ik}(t-t_1)C_{kj}(t_1-t') 
\end{aligned}
\label{eq:mct.mct1}
\end{equation}
with the memory kernel given by 
\begin{equation}
\begin{aligned}
M_{ij}(t)
= 
-\frac{1}{2}
{\cal V}_{i\alpha\beta}C_{\alpha\lambda}(t)C_{\beta\mu}(t)
\left\{ 
{\cal V}_{k \lambda\mu}-2{\cal L}_{k \lambda\mu}
\right\}
C^{-1}_{kj}(0)
.
\end{aligned}
\end{equation}
These equations are almost identical to the conventional mode-coupling
equations\cite{deker1975,bouchaud1996} except for 
$(-2{\cal L}_{k \lambda\mu})$ 
term which appears in the vertex term 
of the memory kernel.
The third term 
on the left hand side of eq.(\ref{eq:mct.mct1}) 
enters naturally to guarantee the time-reversal
symmetry of the correlation function. 
${\cal L}_{ijk}$ is defined by 
\begin{equation}
{\cal L}_{ijk} \equiv 
L^{(0)}_{i\alpha}S^{(3)}_{\alpha jk}
+
L^{(1)}_{i\alpha,j}S^{(2)}_{\alpha k}
+
L^{(1)}_{i\alpha,k}S^{(2)}_{\alpha j}
,
\end{equation}
where we have defined $S^{(2)}_{ij}$ and $S^{(3)}_{ijk}$ by 
\begin{equation}
\begin{aligned}
S^{(2)}_{ij} 
&
\equiv 
\frac{\partial^2 S}{\partial x_{i}\partial x_{j}}\biggl|_{{\bf x}=0}
\hspace*{0.4cm}
\mbox{and}
\hspace*{0.4cm}
S^{(3)}_{ijk} 
\equiv 
\frac{\partial^3 S}{\partial x_{i}\partial x_{j}\partial x_{k}}\Biggl|_{{\bf x}=0}
.
\end{aligned}
\end{equation}
The term $(-2{\cal L}_{k \lambda\mu})$ 
is a new term which originates from
the multiplicative noise. 
Since the density- and velocity-dependent part of the entropy is given
by 
\begin{equation}
S = 
\int\!\!\mbox{d}{\bf r}~ 
\frac{J^2({\bf r})}{2T\rho({\bf r})} 
+ \frac{1}{T}{\cal F},
\end{equation}
$S^{(2)}_{ij}$ and $S^{(3)}_{ijk}$ are given by 
\begin{equation}
\left\{
\begin{aligned}
&
S^{(2)}_{J_{{\bf k}}J_{{\bf k}'}}
= -\frac{1}{m\rho_{0} TV}\delta_{{\bf k},-{\bf k}'}
\\
&
S^{(2)}_{\rho_{{\bf k}}\rho_{{\bf k}'}}
= -\frac{k_{\mbox{\scriptsize B}}}{\rho_{0} S(k)V}\delta_{{\bf k},-{\bf k}'}
\\
&
S^{(3)}_{\rho_{{\bf k}}\rho_{{\bf k}'}\rho_{{\bf k}''}}
= \frac{k_{\mbox{\scriptsize B}}}{\rho_{0}^2V^2}\delta_{{\bf k}+{\bf k}'+{\bf k}'',0}
\\
&
S^{(3)}_{J_{{\bf k}}J_{{\bf k}'}\rho_{{\bf k}''}}
= \frac{1}{m\rho_{0}^2TV^2}\delta_{{\bf k}+{\bf k}'+{\bf k}'',0}
.
\end{aligned}
\right.
\end{equation}
Combining this with eq.(\ref{eq:mct.L}), we find that
${\cal L}_{ijk}=0$ for all elements. 
Therefore, the final expression for the mode-coupling equation 
in the present case
becomes equivalent with the conventional one\cite{gotze1992}. 
Note that the situation will change if we consider different physical
situation:
For example, if we start with the diffusion equation, which is
obtained in the overdamped limit of eq.(\ref{eq:nonlinear}), 
${\cal L}_{ijk}$ plays an essential role\cite{kawasaki1997e}.

Now let us substitute all matrix elements given by
eqs.(\ref{eq:mct.muij}) and
(\ref{eq:mct.Vijk}) to eq.(\ref{eq:mct.mct1}), we have the set of
equations given by 
\begin{equation}
\begin{aligned}
&
\left(\frac{\partial {~}}{\partial t} 
-{\bf k}\cdot{\mbox{\boldmath$\Gamma$}}\cdot
\frac{\partial {~}}{\partial {\bf k} } 
\right)
C_{\rho_{{\bf k}}\rho_{{\bf k}'}}(t)
= 
-ik C_{J_{{\bf k}}\rho_{{\bf k}'}}(t)
\\
&
\left( 
\frac{\partial ~}{\partial t } 
-{\bf k}\cdot{\mbox{\boldmath$\Gamma$}}\cdot
\frac{\partial ~}{\partial {\bf k} } 
-\hat{\bf k}\cdot{\mbox{\boldmath$\Gamma$}}\cdot\hat{\bf k}
\right)
C_{J_{{\bf k}}\rho_{{\bf k}'}}(t)
\\
&
= 
-\frac{ik}{m\beta S(k)} C_{\rho_{{\bf k}}\rho_{{\bf k}'}}(t)
-\frac{\zeta_{0}}{m} C_{J_{{\bf k}}\rho_{{\bf k}'}}(t)
\\
&
-
\frac{1}{m}\int_{0}^{t}\!\!{\mbox{d}} t_{1}~
\sum_{{\bf k}''}
\delta\zeta({\bf k}, {\bf k}'', t-t_1) C_{J_{{\bf k}''}\rho_{{\bf k}'}}(t_1)
,
\end{aligned}
\label{eq:mct.mct2}
\end{equation}
where
\begin{equation}
\sum_{{\bf k}} \equiv \frac{V}{(2\pi)^2}\int\!\!{\mbox{d}}{\bf k}
.
\end{equation}
The memory kernel $\delta\zeta({\bf k}, {\bf k}', t)$ is given by
\begin{equation}
\begin{aligned}
&
\delta\zeta({\bf k}, {\bf k}', t)
\\
&
= \frac{m}{2}\sum_{{\bf q}_1}\sum_{{\bf q}_2}\sum_{{\bf p}_1}\sum_{{\bf p}_2}\sum_{{\bf k}''}
  {\cal V}_{J_{{\bf k}}\rho_{{\bf q}_1}\rho_{{\bf q}_2}}
  C_{\rho_{{\bf q}_1}\rho_{{\bf p}_1}}(t)C_{\rho_{{\bf q}_2}\rho_{{\bf p}_2}}(t)
\\
&
\hspace*{1.0cm}
\times
  {\cal V}_{J_{{\bf k}''}\rho_{{\bf p}_1}\rho_{{\bf p}_2}}
  C_{J_{{\bf k}''}J_{{\bf k}'}}^{-1}(0)
\\
&
= \frac{\beta}{2n V}\sum_{{\bf q}}\sum_{{\bf p}}
  {\cal V}_{J_{{\bf k}}\rho_{{\bf q}}\rho_{{\bf k}-{\bf q}}}
  C_{\rho_{{\bf q}}\rho_{{\bf p}}}(t)C_{\rho_{{\bf k}-{\bf q}}\rho_{-{\bf k}'-{\bf p}}}(t)
\\
&
\hspace*{1.0cm}
\times
  {\cal V}_{J_{-{\bf k}'}\rho_{{\bf p}}\rho_{-{\bf k}'-{\bf p}}}
\\
&
= \frac{1}{2n^2N\beta}\int_{{\bf q}}\int_{{\bf p}}
  V_{{\bf k}}({{\bf q}},{{\bf k}-{\bf q}})
  C_{\rho_{{\bf q}}\rho_{-{\bf p}}}(t)C_{\rho_{{\bf k}-{\bf q}}\rho_{-{\bf k}'+{\bf p}}}(t)
\\
&
\hspace*{1.0cm}
\times
  V_{{\bf k}'}({{\bf p}}, {{\bf k}'-{\bf p}})
,
\end{aligned}
\label{eq:mct.deltazeta}
\end{equation}
where $N$ is the total number of the particles in the system and 
we have defined the vertex function
\begin{equation}
V_{{\bf k}}({{\bf q}},{{\bf k}-{\bf q}})
= 
 \hat{\bf k}\cdot{\bf q} nc(q)
+\hat{\bf k}\cdot ({\bf k}-{\bf q})nc(|{\bf k}-{\bf q}|) 
.
\end{equation}
In the derivation of eq.(\ref{eq:mct.deltazeta}), we 
have used the momentum equal-time correlation function given by 
\begin{equation}
C_{J_{{\bf k}}J_{{\bf k}'}}(0)
= \frac{mnV}{\beta}\delta_{{\bf k},-{\bf k}'}
.
\end{equation}
If the system is in equilibrium, time and space translational
invariance is satisfied and the correlation function becomes 
$C_{\alpha({\bf r})\beta({\bf r}')}(t)
=C_{\alpha({\bf r}-{\bf r}')\beta(\bf 0)}(t)$
or, in ${\bf k}$ space, 
$C_{\alpha_{{\bf k}}\beta_{{\bf k}'}}(t)
=C_{\alpha_{{\bf k}}\beta_{-{\bf k}}}(t)\times\delta_{{\bf k},-{\bf k}'}$.
In the presence of shear, however, 
this should be modified as\cite{onuki1997},
\begin{equation}
C_{\alpha({\bf r})\beta({\bf r}')}(t) 
= 
C_{\alpha({\bf r}-{\bf r}'(t))\beta({\bf 0})}(t) 
,
\label{eq:mct.transinv}
\end{equation}
where we defined the time-dependent position vector by 
\begin{equation}
{\bf r}(t) \equiv \exp[\mbox{\boldmath$\Gamma$} t]\cdot{\bf r} 
= {\bf r} + \dot{\gamma} t y \hat{\bf e}_{x}
,
\label{eq:mct.rt}
\end{equation}
where $\hat{\bf e}_{x}$ is an unit vector oriented along the $x$-axis.
In wave vector space, this is expressed as 
\begin{equation}
\begin{aligned}
C_{\alpha_{\bf k}\beta_{{\bf k}'}}(t) 
&
= 
C_{\alpha_{{\bf k}(t)}\beta_{{\bf -k}}}(t) 
\times \delta_{{\bf k}(t),-{\bf k}'}
\\
&
= 
C_{\alpha_{{\bf k'}(t)}\beta_{{\bf k'}}}(t) 
 \times \delta_{{\bf k},-{\bf k'}(-t)}
\end{aligned}
\label{eq:mct.inv}
\end{equation}
with the time-dependent wave vector defined by 
\begin{equation}
{\bf k}(t) = \exp[{}^{t}\mbox{\boldmath$\Gamma$} t]\cdot{\bf k} 
= {\bf k} + \dot{\gamma} t k_{x}\hat{\bf e}_{y}
,
\label{eq:mct.kt}
\end{equation}
where ${}^{t}\mbox{\boldmath$\Gamma$}$ denotes the transpose of
$\mbox{\boldmath$\Gamma$}$. 
\begin{figure}[t]
\includegraphics[width=0.85\linewidth]{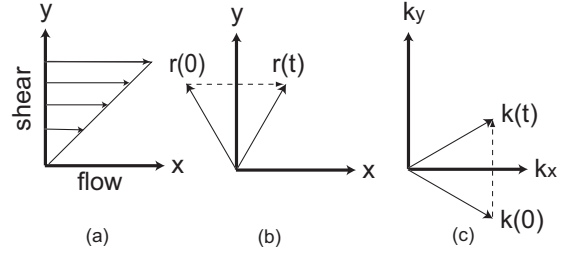}
\caption{
(a) Geometry of shear flow. (b) Shear advection in real space.
(c) Shear advection in Fourier space.
}
\label{flow}
\end{figure}
Figure 1 (b) shows how the shear flow 
advects a positional vector {\bf r} by eq.(\ref{eq:mct.rt}) 
in a time interval of duration $t$.
The corresponding time-dependent wave vector, eq.(\ref{eq:mct.kt}), is 
shown in Fig.1 (c).
Eqs.(\ref{eq:mct.transinv}) and (\ref{eq:mct.inv}) 
state that the fluctuations satisfy
translational invariance in a reference frame flowing with the shear
contours.  
Using this property, eq.(\ref{eq:mct.deltazeta}) becomes
\begin{equation}
\begin{aligned}
&
\delta\zeta({\bf k}, {\bf k}', t)
\\
&
= \frac{1}{2\rho_{0}^2 N\beta}\int_{{\bf q}}\int_{{\bf p}}
  V_{{\bf k}}({{\bf q}},{{\bf k}-{\bf q}})
  C_{\rho_{{\bf q}}\rho_{-{\bf q}(t)}}(t)
  C_{\rho_{{\bf k}-{\bf q}}\rho_{-{\bf k}(t)+{\bf q}(t)}}(t)
\\
&
\hspace*{1.0cm}
\times
  V_{{\bf k}(t)}({{\bf p}(t)}, {{\bf k}(t)-{\bf q}(t)})
  \delta_{{\bf p},{\bf q}(t)}\delta_{{\bf k}',{\bf k}(t)}
\\
&
= \frac{1}{2\rho_{0}^2 NV\beta}\int_{{\bf q}}
  V_{{\bf k}}({{\bf q}},{{\bf k}-{\bf q}})
  C_{\rho_{{\bf q}}\rho_{-{\bf q}(t)}}(t)
  C_{\rho_{{\bf k}-{\bf q}}\rho_{-{\bf k}(t)+{\bf q}(t)}}(t)
\\
&
\hspace*{1.0cm}
\times
  V_{{\bf k}(t)}({{\bf p}(t)}, {{\bf k}(t)-{\bf q}(t)})
  \delta_{{\bf k}',{\bf k}(t)}
\\
&
\equiv
\delta\zeta({\bf k}, t)\times  \delta_{{\bf k}',{\bf k}(t)}
.
\end{aligned}
\label{eq:mct.deltazeta2}
\end{equation}
Introducing the intermediate scattering function by 
\begin{equation}
F({\bf k}, t) 
\equiv \frac{1}{N}
C_{\rho_{{\bf k}(-t)}\rho_{{\bf -k}}}(t)
= \frac{1}{N}
\langle \delta\rho_{{\bf k}(-t)}(t)\delta\rho_{{\bf -k}}(0) \rangle
,
\label{eq:mct.Fkt}
\end{equation}
$\delta\zeta({\bf k}, t)$ can be rewritten as 
\begin{equation}
\begin{aligned}
&
\delta\zeta({\bf k}, t)
= \frac{1}{2\rho_{0}\beta}\int_{{\bf q}}
  V_{{\bf k}}({{\bf q}},{{\bf k}-{\bf q}})
  F({\bf q}(t),t)F({\bf k}(t)-{\bf q}(t),t)
\\
&
\hspace*{1.0cm}
\times
  V_{{\bf k}(t)}({{\bf p}(t)}, {{\bf k}(t)-{\bf q}(t)})
.
\end{aligned}
\label{eq:mct.deltazeta3}
\end{equation}
This is the memory kernel in the presence of the shear. 
This has exactly the same structure as the equilibrium one\cite{gotze1992}
except for the time dependence appearing on the wave vectors. 
The physical interpretation of eq.(\ref{eq:mct.deltazeta3}) is simple: 
The memory kernel has the general structure of two correlation functions 
sandwiched by two vertex functions. 
This means that the interactions of 
two fluctuations with modes ${\bf q}$ and ${\bf k}-{\bf q}$ scatter at a certain
time and then propagate freely in a ``mean field''
and after a time $t$, they recollide and interact. 
Under shear, however, by the time the second interactions take place, 
the fluctuations are streamed away by the flow. 

We also define the cross correlation function
\begin{equation}
H({\bf k}, t) 
\equiv 
\frac{1}{N}
C_{J_{{\bf k}(-t)}\rho_{{\bf -k}}}(t)
=
\frac{1}{N}
\langle J_{{\bf k}(-t)}(t) \delta\rho^{*}_{{\bf k}}(0) \rangle
.
\end{equation}
Then, eq.(\ref{eq:mct.mct2}) can be rewritten as 
\begin{equation}
\begin{aligned}
&
\frac{\mbox{d}F({\bf k}, t) }{\mbox{d}t}
= 
-ik(-t) H({\bf k}, t).
\\
&
\frac{\mbox{d} H({\bf k}, t) }{\mbox{d} t}
= 
-\hat{\bf k}(-t)\cdot\mbox{\boldmath$\Gamma$}\cdot\hat{\bf k}(-t)H({\bf k}, t)
\\
&
- \frac{ik(-t)}{m\beta S(k(-t))}F({\bf k}, t)
-\frac{\zeta_{0}}{m} H({\bf k}, t)
\\
&
-\frac{1}{m}\int_{0}^{t}\!\!\mbox{d} t'~
\delta \zeta({\bf k}(-t),t-t')H({\bf k}, t').
\end{aligned}
\label{eq:mct.mct3}
\end{equation}
Note that in the above equation, the differential operator 
${\bf k}\cdot\mbox{\boldmath$\Gamma$}\cdot\partial/\partial{\bf k}$ 
disappears because
\begin{equation}
\frac{\mbox{d} F({\bf k}, t) }{\mbox{d} t}
= \frac{\partial F({\bf k}, t) }{\partial t}
 -{\bf k}(-t)\cdot\mbox{\boldmath$\Gamma$}
  \cdot\frac{\partial~}{\partial{\bf k}(-t)}F({\bf k}, t).
\label{eq:mct.convectiondelete}
\end{equation}
The memory kernel eq.(\ref{eq:mct.deltazeta3})
appearing in eq.(\ref{eq:mct.mct3}) contains the nonlinear coupling with
the correlation function itself and therefore the equation should be
solved self-consistently.
Indrani and Ramaswamy have derived an equation
similar to eq.(\ref{eq:mct.mct3}) for the velocity correlations of 
a single tagged
particle in a three-dimensional hard sphere colloidal
suspension\cite{indrani1995}.  
They were interested in the relatively low density regime, and did not 
solve the resulting equation self-consistently. 

For colloidal suspensions the relaxation time of the momentum
fluctuations is of the order of $\tau_{m}= m/\zeta_{0}$ and is much
shorter than the relaxation time for density fluctuations which is of
the order of or longer than $\tau_{d}=\sigma^2/D_{0}$.
In other words, we may invoke the overdamped limit. 
Thus, we neglect 
the inertial term ${\mbox{d}} H({\bf k}, t)/{\mbox{d}} t$.
Likewise the first term of the right hand side in the second equation of
eq.(\ref{eq:mct.mct3}) is estimated to be of order of 
$\dot{\gamma}\tau_m$ 
and thus should be very small as long as the P\'{e}clet
number is small. 
Thus, the equation for
the momentum fluctuations may be written as
\begin{equation}
\begin{aligned}
0
= 
&
-\frac{ik(-t)}{m\beta S(k(-t))}F({\bf k}, t)
-\frac{\zeta_{0}}{m}H({\bf k}, t)
\\
&
-\frac{1}{m}\int_{0}^{t}\!\!\mbox{d} t'~
\delta\zeta({\bf k}(-t),t-t')H({\bf k}, t').
\end{aligned}
\label{eq:mct.overdampted}
\end{equation}
Substituting this back into the first equation of 
(\ref{eq:mct.mct3}), 
we arrive at the equation for the intermediate scattering function;
\begin{equation}
\begin{aligned}
\frac{\mbox{d} F({\bf k}, t) }{\mbox{d}t}
= 
&
-\frac{D_{0}k(-t)^2}{S(k(-t))}F({\bf k}, t)
\\
&
-\int_{0}^{t}\!\!\mbox{d} t'~
  M({\bf k}(-t),t-t')
  \frac{\mbox{d} F({\bf k}, t') }{\mbox{d}t'},
\end{aligned}
\label{eq:mct.mctfinal}
\end{equation}
where the memory kernel is given by 
\begin{equation}
\begin{aligned}
&
M({\bf k},t)
= 
\frac{D_{0}}{2\rho_{0}}
\frac{k}{k(t)}
\int_{{\bf q}}
V_{{\bf k}}({\bf q}, {\bf k}-{\bf q})
V_{\bf k(t)}({\bf q}(t),{\bf k}(t)-{\bf q}(t))
\\
&
\times
F({\bf k}(t)-{\bf q}(t), t)F({\bf q}(t), t).
\end{aligned}
\label{eq:mct.Mkt}
\end{equation}
Eqs.(\ref{eq:mct.mctfinal}) and (\ref{eq:mct.Mkt}) 
are the final mode-coupling expressions.
We have started from the equation for both the density and the momentum
fields and proceeded via the standard mode-coupling approach, taking
 the overdamped limit at the end. 
It should be possible to derive the same result 
starting from the diffusion equation with an interaction term which is
obtained by taking the overdamped limit in eq.(\ref{eq:nonlinear}). 
In order to arrive at the same result, however, one needs to use 
a different 
resummation scheme that involves the 
irreducible projection operator\cite{kawasaki1995c}. 

Eq.(\ref{eq:mct.mctfinal}) is a non-linear integro-differential
equation that can be solved numerically. 
An efficient numerical routine to solve the mode-coupling equation is
elucidated in Ref.\cite{fuchs1991}. 
Since our
 equation is not isotropic in ${\bf k}$ due to the presence of shear, 
the numerics are more involved than in the equilibrium case. 
We shall solve the equation by dividing the lower half plane into 
an $N_{k}\times (N_{k}-1)/2$ grid, where $N_{k}$ is the grid number
which 
we have chosen to be an odd number.
We do not need to consider the upper half plane because it is a mirror
image of the lower one. 
$k_x$, for example, is discretized as 
$k_{x,0}=-k_{c}, k_{x,1}=-k_{c}+\delta_k, 
\cdots, k_{x,N_k}= -k_c+N_k\delta_{k}=k_{c}$, 
where $\delta_{k} = 2k_{c}/N_{k}$
is the grid size and
$k_{c}$ is a cut-off wave vector.
Any value outside the boundary, $|k_x|,~|k_y| > k_c$, 
is replaced by the value at the boundary. 

\section{Simulation method}
\renewcommand{\theequation}{\Roman{section}.\arabic{equation}}  
\setcounter{equation}{0}

To prevent crystallization and obtain stable amorphous states
via molecular dynamics (MD) simulations, 
we choose a model two-dimensional system composed of two different particle species
$1$ and $2$, which interact via the soft-core potential
\begin{equation}
v_{ab}(r)=\epsilon (\sigma_{ab}/r)^{12}
\end{equation}
with $\sigma_{ab}=(\sigma_{a}+\sigma_{b})/2$,
where $r$ is the distance between two particles,
and $a,b$ denote particle species $(\in 1,2)$.
We take the mass ratio to be $m_{2}/m_{1}=2$, the size ratio 
to be $\sigma_{2}/\sigma_{1}=1.4$, and the number of particles
$N=N_1+N_2$, $N_{1}=N_{2}=5000$.
Simulations are performed in the presence and absence of 
shear flow keeping the particle density and the temperature
fixed at $n=n_1+n_2=0.8/\sigma_{1}^{2}$ ($n_1=N_1/V$, $n_2=N_2/V$) and 
$k_BT=0.526\epsilon$, respectively.
Space and time are measured in units of $\sigma_1$ and
$\tau_0=({m_{1}\sigma_{1}^{2}/\epsilon})^{1/2}$.
The size of the unit cell is $L=118 \sigma_{1}$.
In the absence of shear, we impose microcanonical conditions
and integrate Newton's equations of motion
\begin{equation}
\frac{d{\bf r}^a_i}{dt}  =\frac{ {\bf p}^a_i}{m_a},  \quad 
\frac{d{\bf p}^a_i}{dt}  ={\bf f}^a_i
\label{eq:2.5}
\end{equation}
after very long equilibration periods so that no appreciable 
aging (slow equilibration) effect is detected in various quantities 
such as the pressure or in various time correlation functions.
Here, ${\bf r}^a_i=(r^a_{xi},r^a_{yi})$ and 
${\bf p}^a_i=(p^a_{xi},p^a_{yi})$ denote 
the position and the momentum of the $i$-th particle of the 
species $a$, and ${\bf f}^a_i$ is the force acting on the $i$-th
particle of species $a$.
In the presence of shear, by defining the momentum
${\bf p'}^{a}_i={\bf p}^a_i - m_a \dot{\gamma} r^a_{yi}\hat{\bf e}_x$
(the momentum deviations relative to mean Couette flow),
and using the Lee-Edwards boundary condition, 
we integrate the so-called 
SLLOD equations of motion so that 
the temperature 
$k_BT$ $(\equiv N^{-1}{\sum_a\sum_{i}({{\bf p'}}^a_i)^2}/m_a$) 
is kept at a desired value using
a Gaussian constraint thermostat
to eliminate viscous heating effects\cite{Evans}.
The system remains at rest for $t < 0$ for a long
equilibration time and is then sheared for  $t \ge 0$.
Data for analysis has been taken and accumulated 
in steady states which can be realized after transient waiting periods.  

We shall calculate the incoherent and the coherent parts of the 
scattering function for the binary mixture by using the definitions 
\cite{fskt}
\begin{equation}
F_{s}({\bf k},t)=\frac{1}{N_a}{\bigg\langle}\sum_{i=1}^{N_a}
\mbox{\large e}^{[-i\{{\bf k}(-t)\cdot{\bf r}^a_i(t)-{\bf k}\cdot{\bf r}^a_i(0)\}]}{\bigg\rangle}
\label{eq:fskt}
\end{equation}
and 
\begin{equation}
F_{ab}({\bf k},t)=\frac{1}{N}{\bigg\langle}
\sum_{i=1}^{N_a}
\mbox{\large e}^{[-i{\bf k}(-t)\cdot{\bf r}^a_i(t)]}
\sum_{j=1}^{N_b}
\mbox{\large e}^{[i{\bf k}\cdot{\bf r}^b_j(0)]}
{\bigg\rangle},
\label{eq:fkt}
\end{equation}
respectively with $a,b\in 1,2$.
The $\alpha$ relaxation time $\tau_\alpha$ of
the present mixture, which is defined by
\begin{equation}
F_{11}({\bf k}_0,\tau_\alpha)\simeq F_{s}({\bf k}_0,\tau_\alpha)=\mbox{\large e}^{-1}
,
\label{eq:sim.taualphadef}
\end{equation} 
is equal to $\tau_{\alpha}\simeq 1800$ time units
in the quiescent state for $|{\bf k}_0|=2\pi/\sigma_1$.

\section{Results}
\renewcommand{\theequation}{\Roman{section}.\arabic{equation}}  
\setcounter{equation}{0}

\subsection{Microscopic Structure}
\label{subsec:Sk}

The partial static structure factors $S_{ab}({\bf k})$ are defined as
\begin{equation}
S_{ab}({\bf k})=
\int{\mbox{d}}{\bf r}~
\mbox{\large e}^{i{\bf k}\cdot{\bf r}}
\langle{\hat{n}}_{a}({\bf r}){\hat{n}}_{b}({\bf 0})\rangle,
\label{eq:3.6}
\end{equation}
where
\begin{equation}
\hat{n}_{a}({\bf r})=
\sum_{j}^{N_a}\delta ({\bf r}-{\bf r}^a_{j}) \qquad
 (a=1,2)
\label{eq:2.8}
\end{equation}
is the local number density of the species $a$.
Note the dimensionless wave vector ${\bf k}$ is
measured in units of $\sigma_1^{-1}$.
For a binary mixture there are three combinations 
of partial structure factors,
$S_{11}({\bf k})$, $S_{22}({\bf k})$, and $S_{12}({\bf k})$. 
They are plotted in Fig.2 (a) 
in the quiescent state after taking 
an angular average over ${\bf k}$.
A density variable representing the degree of particle packing,
corresponding to the density of an effective one component system, 
can be defined for the present binary system by
\begin{equation}
\hat{\rho}_{\mbox{\scriptsize eff}}({\bf r})= \sigma_1^2 \hat{n}_1({\bf r})
+ \sigma_2^2\hat{n}_2({\bf r}).
\label{eq:reff}
\end{equation}
The corresponding dimensionless structure factor is given by
\begin{equation}
\begin{aligned}
&
S_{\rho\rho}({\bf k})
=  \sigma_1^{-4} \int{\mbox{d}}{\bf r}~
\mbox{\large e}^{i{\bf k}\cdot{\bf r}}
\langle
       \delta\hat{\rho}_{\mbox{\scriptsize eff}}({\bf r})
       \delta\hat{\rho}_{\mbox{\scriptsize eff}}({\bf 0})
\rangle
\\
&
=n_1^2S_{11}({\bf k})+n_2^2(\sigma_2/\sigma_1)^4S_{22}({\bf k})
+2n_1 n_2(\sigma_2/\sigma_1)^2S_{12}({\bf k}),
\end{aligned}
\label{eq:skrr}
\end{equation}
where $\delta\hat{\rho}_{\mbox{\scriptsize eff}}=\hat{\rho}_{\mbox{\scriptsize eff}}-
\langle\hat{\rho}_{\mbox{\scriptsize eff}}\rangle$.  
One can see from Fig.2 (b) 
that $S_{\rho\rho}(k)$ 
has a pronounced peak at  $k \simeq 5.8 $ and becomes
very small ($\sim 0.01)$ at smaller $k$
demonstrating that our system is highly
incompressible at long wavelengths.
Because $S_{\rho\rho}(k)$ behaves quite similarly to $S(k)$ of 
one component systems, we examine space-time correlations 
in $\hat{\rho}_{\mbox{\scriptsize eff}}({\bf r})$
rather than those in the partial number density $\hat{n}_a({\bf r})$ 
for the present binary system.
The usage of $\hat{\rho}_{\mbox{\scriptsize eff}}({\bf r})$ makes comparisons of our 
simulation data with the mode-coupling theory developed for 
a one component system more meaningful.
\begin{figure}[htb]
\includegraphics[scale=0.4,angle=0]{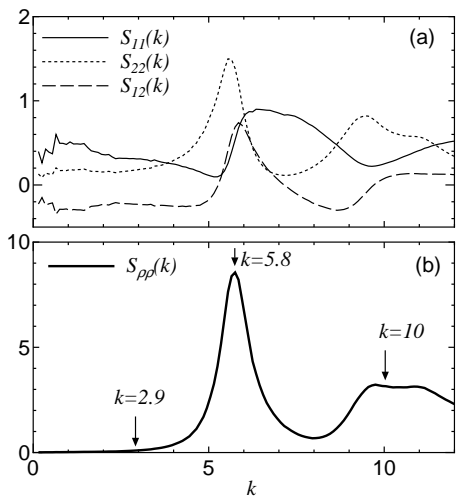} 
\caption{Partial structure factors $S_{ab}(k)$ in (a) and 
$S_{\rho\rho}(k)$ in (b) defined by eq.(\ref{eq:skrr}) 
for the present binary mixture. }
\label{fig:Sktotal}
\end{figure}

We next examine the anisotropy in the static structure factor 
$S_{\rho\rho}({\bf k})$ in the presence of shear flow.
Figures 3--6 show $S_{\rho\rho}({\bf k})$
plotted on a two-dimensional $k_x-k_y$ plane (upper part) 
and the angular averaged curves (lower part) within the regions (a)-(d)
obtained at $\dot{\gamma}=10^{-4}$, $10^{-3}$, $10^{-2}$, 
and $10^{-1}$, respectively (in units of $\tau_0^{-1}$).
One sees that, at the lowest shear rate ($\dot{\gamma}=10^{-4}$), 
the shear distortion is negligible but at higher shear rate the distortion
becomes prominent. 
At $\dot{\gamma}=10^{-3}$, at all regions except for the region (d), 
the peak heights of $S_{\rho\rho}({\bf k})$ 
start decreasing. 
This is contrary to the estimate from the linear response theory
given by eq.(\ref{eq:mct.Sk}) 
which predicts no distortion in the region (a) and
(c) but distortion to the higher, (b), and lower, (d), peaks. 
This seems to indicate that the nonlinear effects due to the shear 
become important. 
Ronis has explored the higher shear regime for the structure factor
of hard sphere colloidal suspensions and concluded that at higher
shear, the peak should be always lower than the equilibrium value 
together with a shift that depends on the direction\cite{ronis1984}. 
Figure 6 shows that peaks in all directions have been 
lowered and the peak with maximal distortion (region (b)) is shifted
to the lower wave vectors while the opposite is true for the shift of
the region with minimal distortion 
(region (d)). 
The qualitative agreement with Ronis' theory
is good but it is not clear that
our results  can be explained by a simple two-body theory such as that
of Ronis. 
Recently Szamel has analyzed $S(k)$ for hard sphere
colloidal suspensions up to the linear order in $\dot{\gamma}$\cite{szamel2001}. 
He took the three-body correlations into account and found 
quantitative agreement with the shear viscosity evaluated using 
$S(k)$. 
\begin{figure}[htb]
\includegraphics[scale=0.38,angle=0]{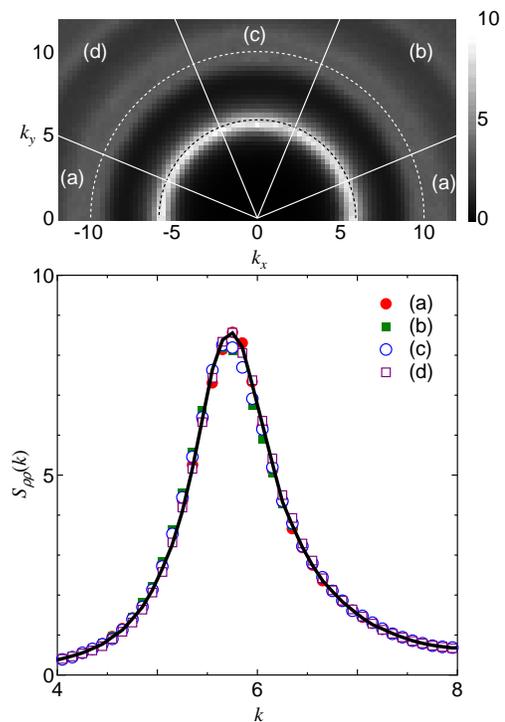} 
\caption{$S_{\rho\rho}(k)$ for $\dot{\gamma}=10^{-4}$.
The solid line is $S_{\rho\rho}(k)$ at equilibrium. 
Dots represents those observed in the region indicated in the
$(k_x,~k_y)$ plane above.}
\label{fig:Sk10E-4}
\end{figure}
\begin{figure}[htb]
\includegraphics[scale=0.38,angle=0]{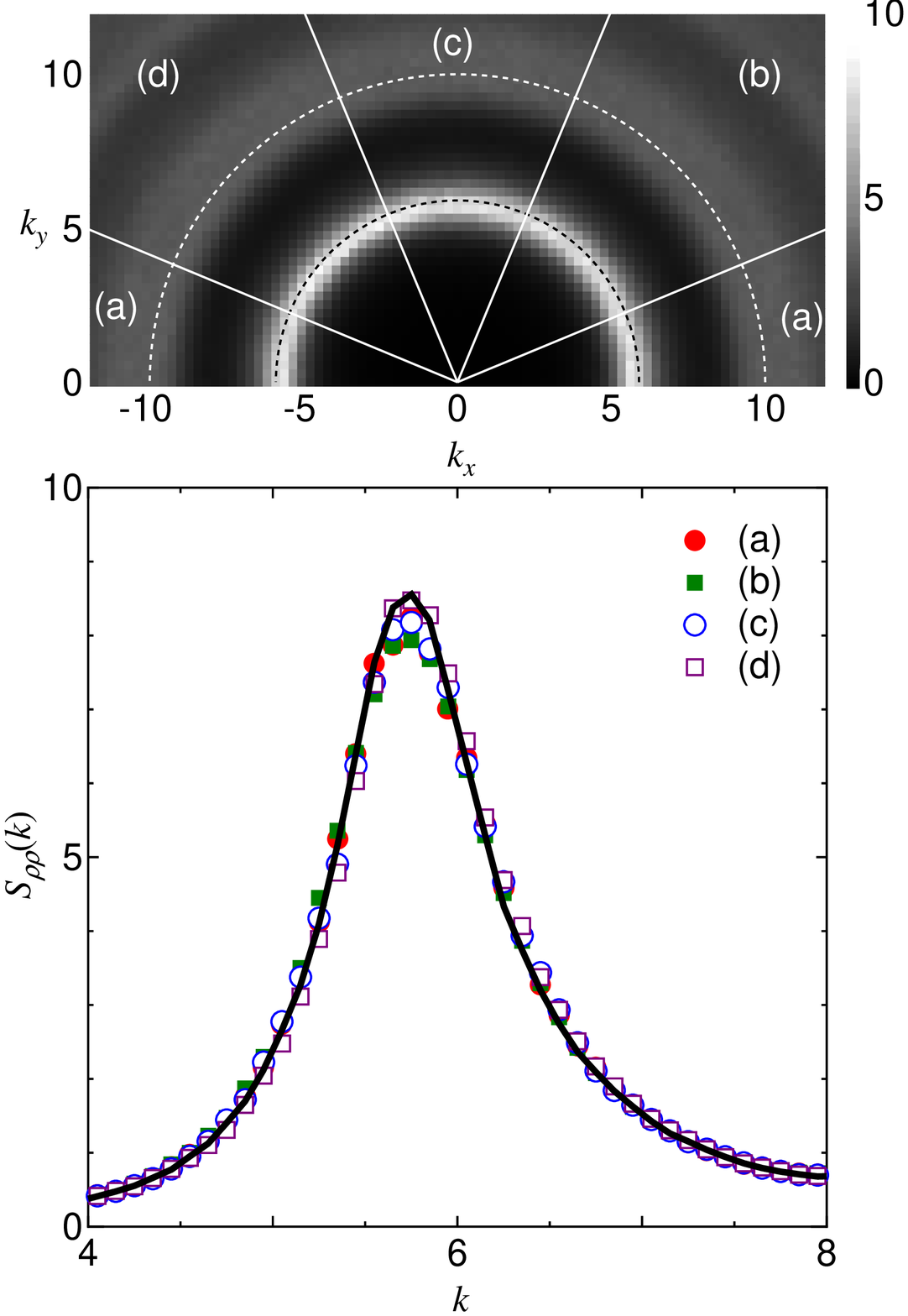} 
\caption{$S_{\rho\rho}(k)$ for $\dot{\gamma}=10^{-3}$.
All symbols are as in Fig.\ref{fig:Sk10E-4}}
\label{fig:Sk10E-3}
\end{figure}
\begin{figure}[htb]
\includegraphics[scale=0.38,angle=0]{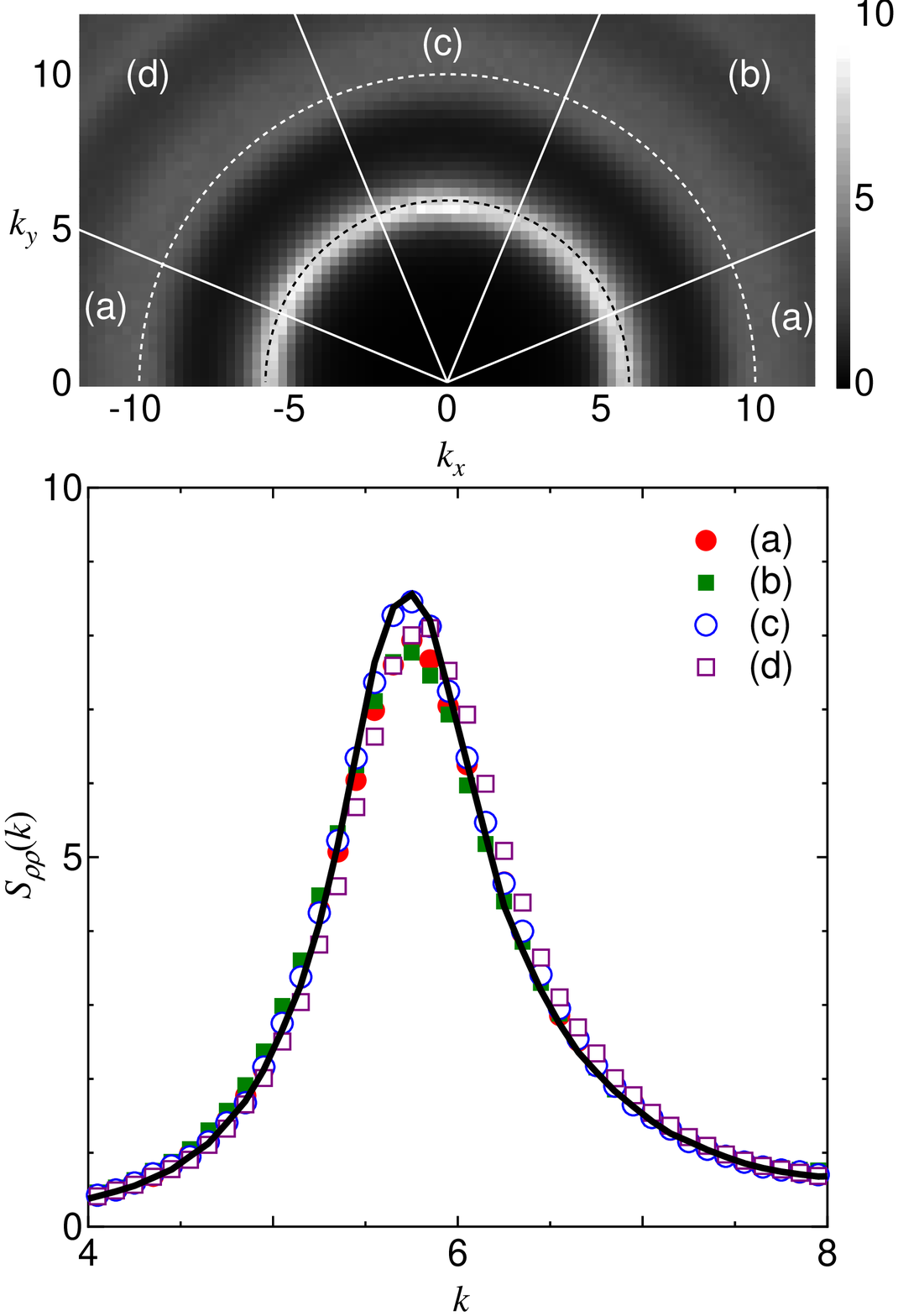} 
\caption{$S_{\rho\rho}(k)$ for $\dot{\gamma}=10^{-2}$.
All symbols are as in Fig.\ref{fig:Sk10E-4}}
\label{fig:Sk10E-2}
\end{figure}
\begin{figure}[htb]
\includegraphics[scale=0.38,angle=0]{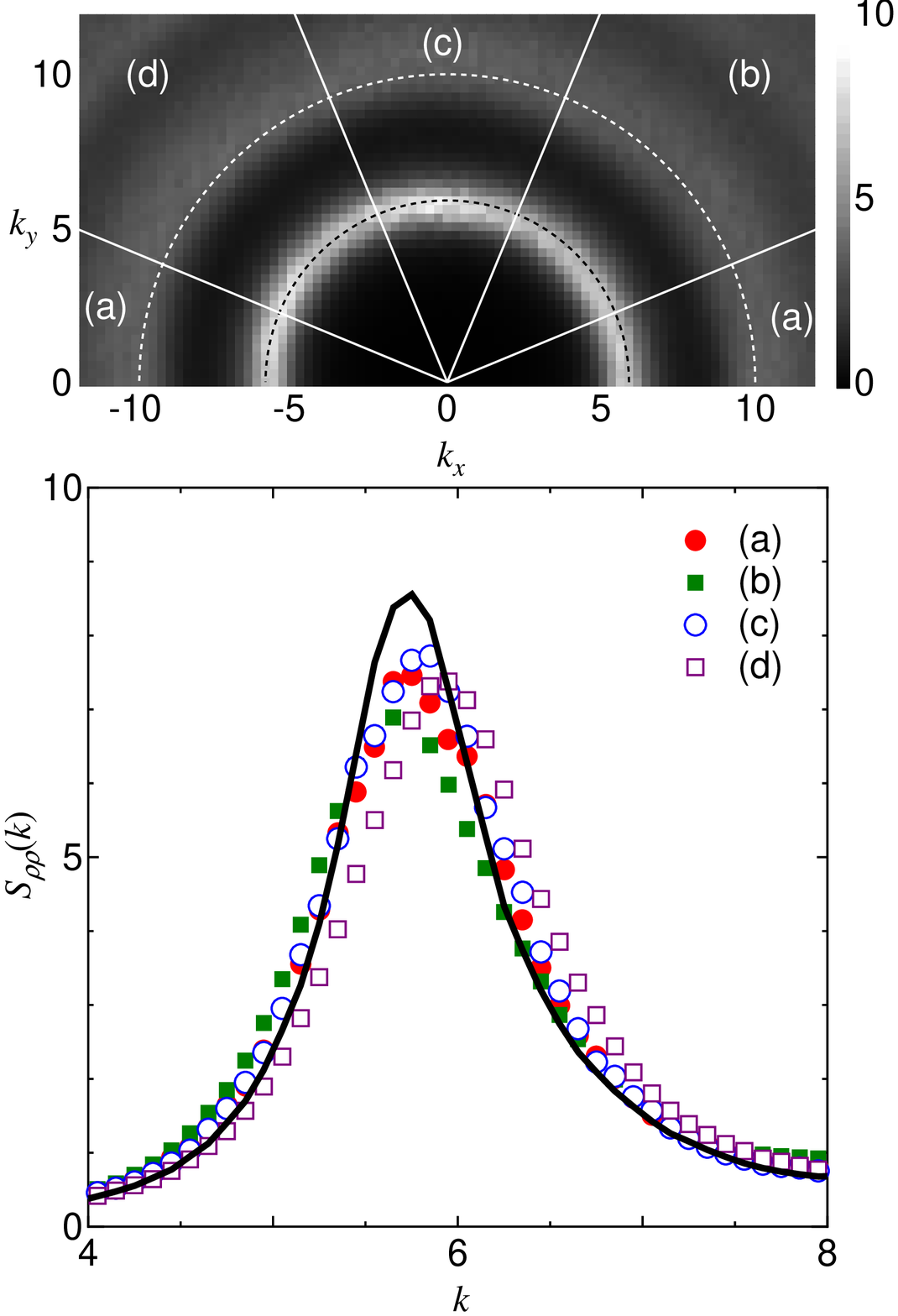} 
\caption{$S_{\rho\rho}(k)$ for $\dot{\gamma}=10^{-1}$.
All symbols are as in Fig.\ref{fig:Sk10E-4}}
\label{fig:Sk10E-1}
\end{figure}

It should be noted that because our system is a liquid, 
we cannot directly refer to 
the P\'{e}clet number 
since the bare diffusion coefficient $D_{0}$ does not exist. 
Therefore, a direct and quantitative comparison with the theories
discussed above 
is not possible. 
However, estimates from the relaxation times self-diffusion coefficient 
evaluated in
Ref.\cite{yamamoto1998c} allow us to estimate a 
P\'{e}clet number in the range 
between  $10^{-1}$ and $10^2$, which corresponds to the highest
shear rates explored in the theoretical analysis of this paper. 

\subsection{Intermediate Scattering Function: Numerical Results}
\label{subsec:resutls.sim}

Here, we examine the dynamics of the local density variable 
$\hat{\rho}_{\mbox{\scriptsize eff}}({\bf r},t)$.
To this end, we defined the intermediate scattering function
\begin{equation}
\begin{aligned}
F_{\rho\rho}({\bf k},t)
=
&
n_1^2F_{11}({\bf k},t)+n_2^2(\sigma_2/\sigma_1)^4F_{22}({\bf k},t)
\\
&
+2n_1 n_2(\sigma_2/\sigma_1)^2F_{12}({\bf k},t)
\end{aligned}
\label{eq:3.3}
\end{equation}
by taking a linear combination of the partial scattering functions
defined in eq.(\ref{eq:fkt}).
Note that $F_{\rho\rho}({\bf k},0)=S_{\rho\rho}({\bf k})$ by definition.
To investigate anisotropy in the scattering function $F_{\rho\rho}({\bf k},t)$, 
the wave vector ${\bf k}$ is taken in four different 
directions ${\bf k}_{10}$, ${\bf k}_{11}$, ${\bf k}_{01}$, 
and ${\bf k}_{-11}$, where
\begin{equation}
{\bf k}_{\mu\nu}=\frac{k}{\sqrt{\mu^2+\nu^2}}(\mu\hat{\bf e}_x+\nu\hat{\bf e}_y)
\label{eq:resut.B.kmunu}
\end{equation}
and $\mu,\nu\in 0,1$ as shown in Fig.7.
The wave vector $k$ (in reduced units) 
is taken to be $2.9$, $5.8$, and $10$ 
(see also Fig.2 (b))
Because we use the Lee-Edwards periodic boundary condition,
the available wave vectors in our simulations should be given by
\begin{equation}
{\bf k}=\frac{2\pi}{L}(n{\hat{\bf e}_x},(m-nD_x){\hat{\bf e}_y}),
\end{equation}
where $n$ and $m$ are integers, $D_x=L\dot{\gamma}t$ is the 
difference in $x$-coordinate between the top and bottom cells
as depicted in Fig.6.5 of Ref.\cite{Evans}.
To suppress statistical errors, we sample about 80
available wave vectors around ${\bf k}_{\mu\nu}$ and calculate
$F_{\rho\rho}({\bf k},t)$ using eqs.(\ref{eq:3.3}) and (\ref{eq:fkt}),
and then we average $F_{\rho\rho}({\bf k},t)$ over sampled wave vectors.
The sampled wave vectors are shown as the spots in Fig.7.
\begin{figure}[htb]
\includegraphics[scale=0.43,angle=0]{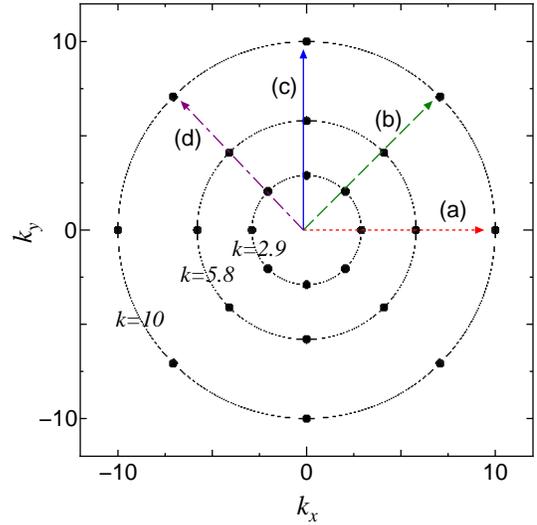} 
\caption{Sampled wave vectors.}
\label{fig:kvec}
\end{figure}

Figures 8, 9, and 10
show $F_{\rho\rho}({\bf k},t)/S_{\rho\rho}({\bf k})$
for $k=2.9$, $5.8$, and $10$, respectively.
We have observed that all of the partial scattering functions 
$F_{11}({\bf k}, t)$, $F_{12}({\bf k}, t)$, and 
$F_{22}({\bf k}, t)$ behave in a similar manner as 
$F_{\rho\rho}({\bf k},t)$, which demonstrates that the effective 
single component scattering function typifies the dynamics of the whole
system. 
Several features are noticeable. 
First, the quantitative trends as a function of $k$ are similar for
different values of $\dot{\gamma}$. 
Secondly,
shear drastically accelerates microscopic structural relaxation in
the supercooled state. The structural relaxation time $\tau_\alpha$
decreases strongly with increasing shear rate as $\tau_\alpha\sim
\dot{\gamma}^{-\nu}$ with $\nu \simeq 1$.
Lastly, the acceleration in the dynamics due to shear occurs almost 
isotropically. 
We observed surprisingly small anisotropy in the
scattering functions even under extremely strong shear, 
$\dot{\gamma}\tau_\alpha\simeq 10^3$.
A similar isotropy in the tagged particle motions has already 
been reported in Ref.\cite{yamamoto1998c}.
The observed isotropy is more surprising than that observed 
in single particle quantities. 
In particular, the fact that different particles  labels are correlated
in the collective quantity defined in eq.(\ref{eq:fkt}) means that a
simple transformation to a frame moving with the shear flow cannot
completely removed the directional character of the shear. 
Our results provide {\it post facto} justification for the isotropic
approximation of Ref.\cite{fuchs2002}. 
This simplicity in the dynamics is quite different from behavior 
of other complex fluids such as critical fluids or polymers, 
where the dynamics become noticeably 
anisotropic in the presence of shear flow.
\begin{figure}[htb]
\includegraphics[scale=0.43,angle=0]{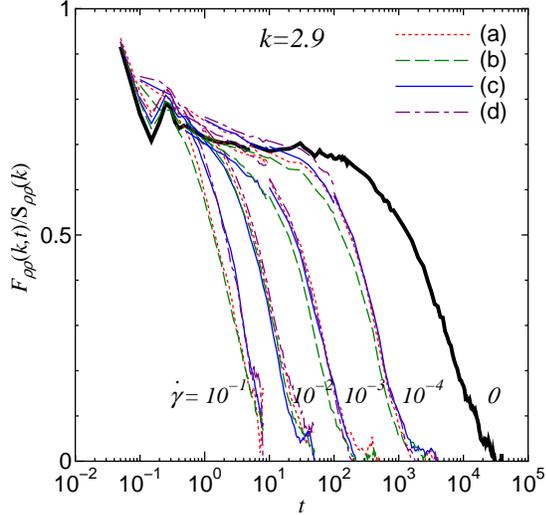} 
\caption{
$F({\bf k}, t)/S({\bf k})$ at $k\sigma_{1}=2.8$ for various shear
rates and at the different observing points 
(a) ${\bf k}_{10}$, (b) ${\bf k}_{11}$, (c) ${\bf k}_{01}$, 
and (d) ${\bf k}_{-11}$ as explained in Fig.{\ref{fig:kvec}}.
}
\label{fig:MDFktk29}
\end{figure}

\begin{figure}[htb]
\includegraphics[scale=0.43,angle=0]{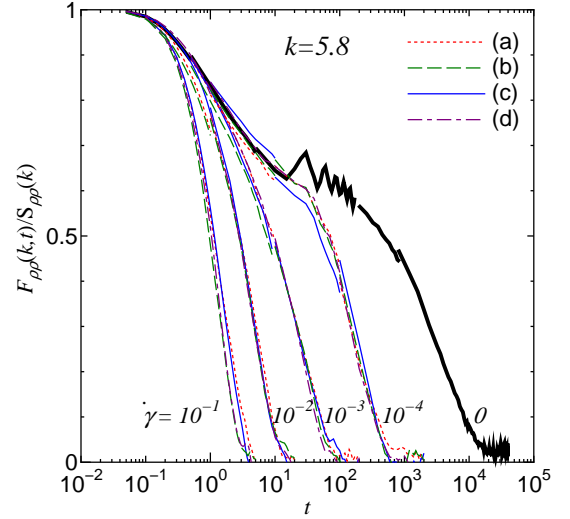} 
\caption{$F({\bf k}, t)/S({\bf k})$ at $k\sigma_{1}=5.8$.
All symbols are as in Fig.\ref{fig:MDFktk29}}
\label{fig:MDFktk58}
\end{figure}

\begin{figure}[ht]
\includegraphics[scale=0.43,angle=0]{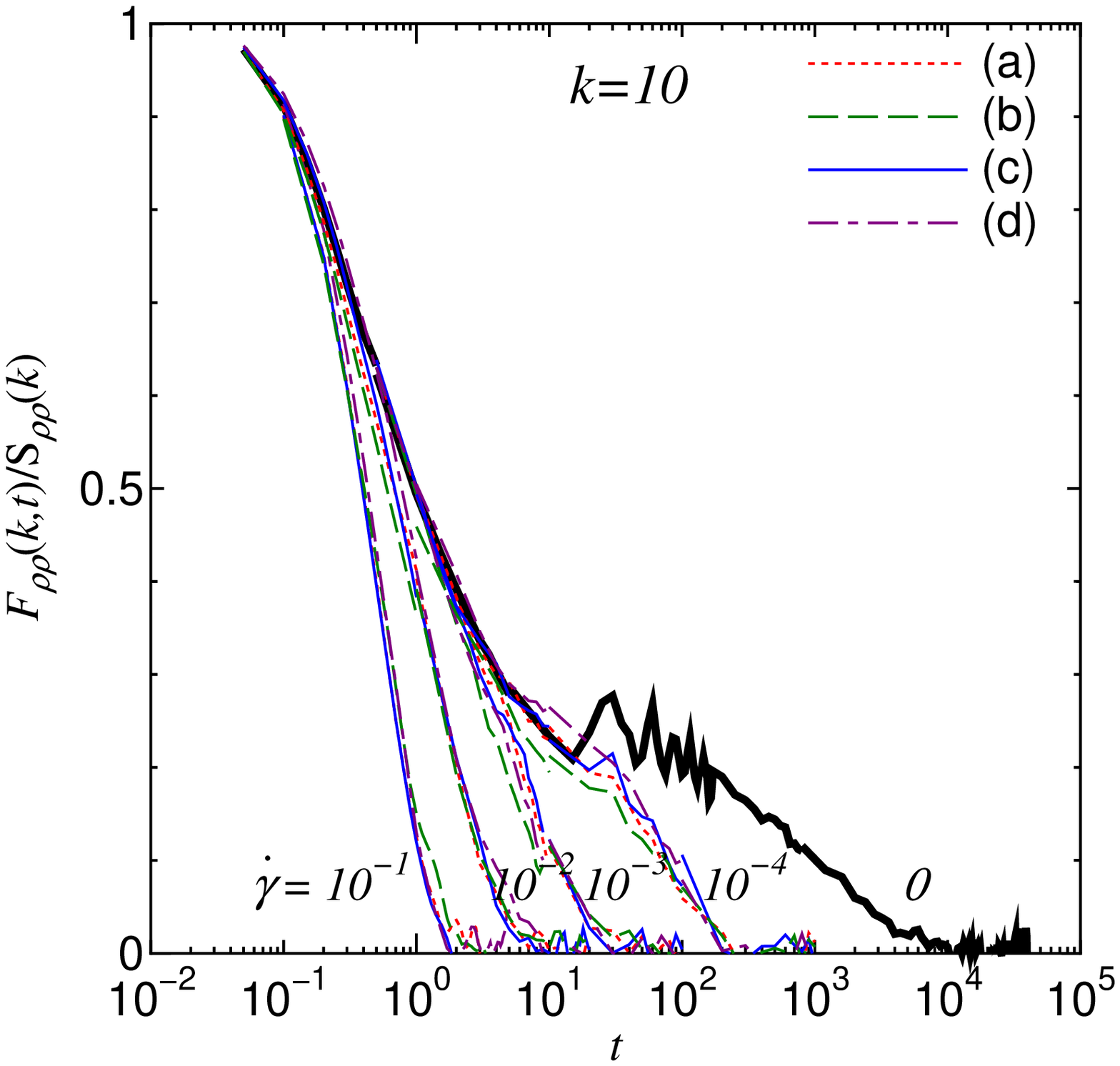} 
\caption{$F({\bf k}, t)/S({\bf k})$ at $k\sigma_{1}=10$.
All symbols are as in Fig.\ref{fig:MDFktk29}}
\label{fig:MDFktk100}
\end{figure}

\subsection{Intermediate Scattering Function: MCT Results}
\label{subsec:resutls.mct}

\begin{figure}[ht]
\includegraphics[scale=0.6,angle=0]{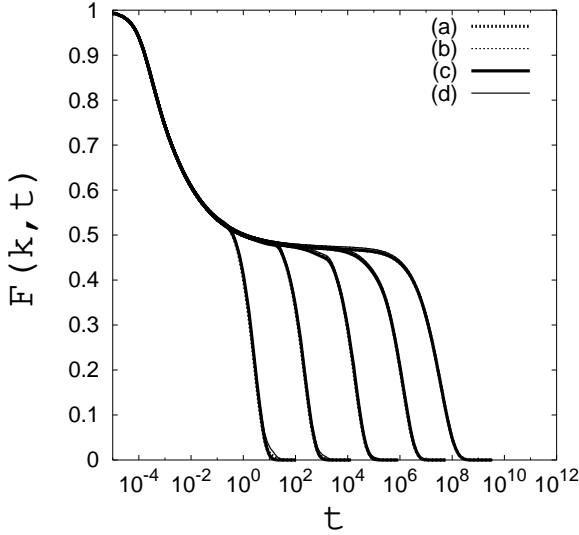} 
\caption{
$F({\bf k}, t)/S(k)$ for 
$k\sigma=3$ 
for different observing points and for 
various shear rates. 
From the right to the left, 
$\mbox{Pe} =10^{-10}$, $10^{-7}$, 
$10^{-5}$, $10^{-3}$, and $10^{-1}$.
The thick dotted lines are at (a) in Fig.\ref{fig:kvec}.
The thin  dotted lines are at (b).
The thick solid  lines are at (c).
The thin solid  lines are at (d).
The density is $\phi=  \phi_c \times (1 - 10^{-4})$.
The time $t$ is scaled by $\sigma^2/D_{0}$. 
}
\label{fig:mctk30}
\end{figure}
\begin{figure}[ht]
\includegraphics[scale=0.6,angle=0]{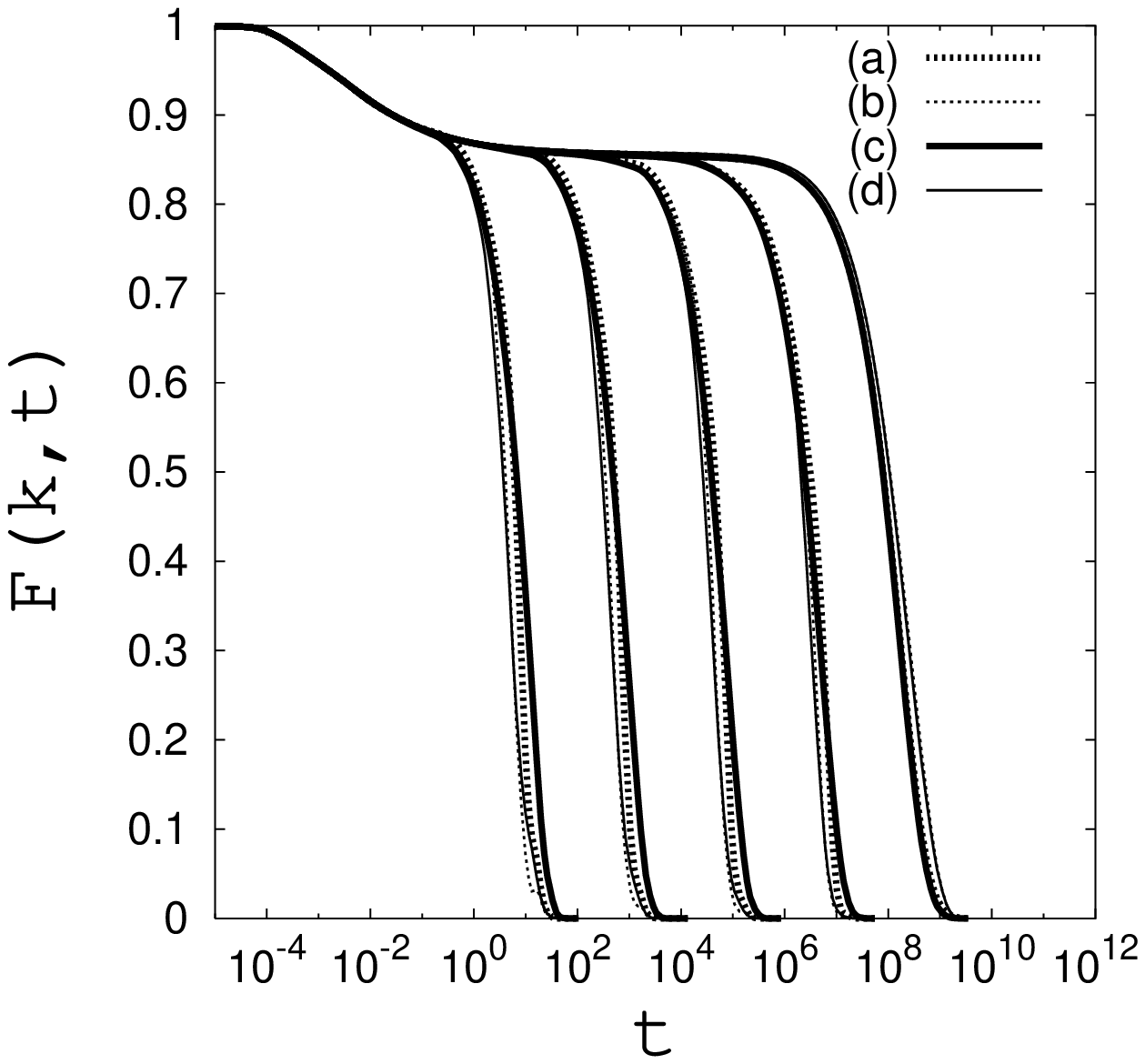} 
\caption{$F({\bf k}, t)/S(k)$ for $k\sigma=6.7$.
All symbols are as in Fig.\ref{fig:mctk30}}
\label{fig:mctk67}
\end{figure}
\begin{figure}[ht]
\includegraphics[scale=0.6,angle=0]{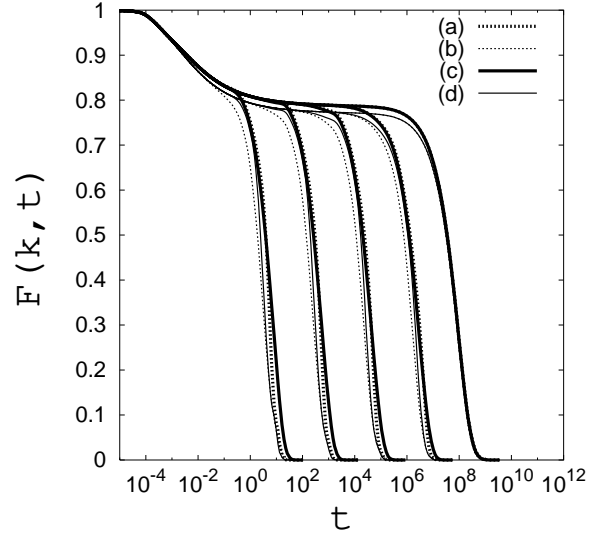} 
\caption{$F({\bf k}, t)/S(k)$ for $k\sigma=12$.
All symbols are as in Fig.\ref{fig:mctk30}} 
\label{fig:mctk120}
\end{figure}
We evaluate $F(k, t)$ for the two-dimensional colloidal suspension
theoretically using 
eq.(\ref{eq:mct.mctfinal}) with eq.(\ref{eq:mct.Mkt}). 
Following the procedure explained in Section \ref{sec:mct}, 
we have solved the equations eq.(\ref{eq:mct.mctfinal}) self-consistently. 
For the static correlation function $c(k)$ and $S(k)$, 
the analytic expressions derived by Baus {\it et al.}
were used (see Appendix A)\cite{baus1987}. 
The number of grid points was chosen to be $N_k=55$. 
We have also calculated solution of eq.(\ref{eq:mct.mctfinal})
for 
the larger grid sizes up to $N_k=101$ but qualitative
differences between results obtained with different grid numbers were not
noticeable. 
We chose $k_c\sigma = 10\pi$ as the 
cut-off wave vector. 
The best estimate for the transition density is 
$\phi_c=0.72574$ for $N_{k}=670$ at equilibrium, where
$\phi=\pi \sigma^2 \rho_{0}/4$ is the volume fraction. 
For $N_k=55$, a higher value, 
$\phi_c=0.76645$ is obtained. 
Figures 11--13
show the behavior of $F({\bf k}, t)$ renormalized by its initial value 
$S(k)$
for $\phi=  \phi_c \times (1 - 10^{-4})$ for various 
shear rates from Pe=$10^{-10}$ to $10^{-1}$, where 
Pe is the P\'{e}clet number defined by 
Pe$=\dot{\gamma}\sigma^2/D_0$.   
The wave vectors were chosen to be 
$k\sigma=3.0$ (Fig.11),
$k\sigma=6.7$ (Fig.12),
and
$k\sigma=12.0$ (Fig.13). 
$k\sigma=6.7$ is close to the position of the first peak of $S(k)$. 
At each wave vector, we have observed $F({\bf k},t)/S(k)$ for four directions
denoted by 
(a)-(d) in Fig.7 and defined by eq.(\ref{eq:resut.B.kmunu}).
For shear rates smaller than $\mbox{Pe} = 10^{-10}$ 
no effect of shear is observed.
For Pe $\geq 10^{-10}$, 
we observe a large reduction of relaxation times
due to shear.
We define the structural relaxation time, $\tau_{\alpha}$, 
as in eq.(\ref{eq:sim.taualphadef}). 
We find that, although the amplitudes of relaxation 
defer depending on $k$, 
the shear dependence of the relaxation time and its shear dependence is
almost independent of $k$. 
The shear rate dependence of the relaxation time is given by 
$\tau_{\alpha}(\dot{\gamma})\sim \dot{\gamma}^{-1}$,  
consistent with the simulation results discussed in the previous
subsection.    
The four curves for a fixed shear rate but for different 
wave vector directions exhibit almost perfect isotropy, 
in qualitative agreement with the behavior observed in the simulations. 
Thus the apparent anisotropy of the vertex function in
eq.(\ref{eq:mct.Mkt}) provides virtually no anisotropic scattering. 
For $k\sigma =3$, we see perfectly isotropic scattering.
For $k\sigma=6.7$ and 12.0, 
the curves show small anisotropy that is still consistent with the
simulation results.  
Note that the differences shown in the 
plateau value for $k\sigma=12.0$
is an artifact due to our use of a square grid
which produces an error in the radial distances depending on 
direction. 
The differences are noticeable but small in (b) and (d) for
$k\sigma=12.0$. 
In these calculations, we have used the static structure factor 
at equilibrium and the distortion of the structure due to the shear
was neglected. 
However, as shown in Figures 3--6, 
the shape of $S(k)$ becomes weakly anisotropic under shear. 
In order to see if the anisotropy of the structures affects 
the dynamics, we have implemented the same MCT calculation 
using $S(k)$ and $nc(k)$ with 
an anisotropic sinusoidal modulation mimicking those observed in the
simulations.  
We found that, as long as modulations are small, 
no qualitative change was observed and the dynamics was still isotropic. 

It is surprising that, although the perturbation 
is highly
anisotropic, the dynamics of fluctuations are almost isotropic. 
The reason for the isotropic nature of fluctuations may be 
understood as follows: 
The shear flow perturbs and
randomizes the phase of coupling between different modes.  
This perturbation dissipates the cage that transiently immobilizes
particles.  
Mathematically, this is reflected through the time
dependence of the vertex.
This ``phase randomization'' occurs irrespective of the direction of the
wave vector, which results in the essentially isotropic behavior of 
relaxation. 
This mechanism is different from 
that of many complex fluids and of dynamics near a critical point
under shear, in which anisotropic distortion of 
the fluctuations at small wave vectors by shear 
plays an essential role\cite{onuki1997}.

\subsection{Viscosity}

The shear-dependent viscosity $\eta(\dot{\gamma})$ 
is evaluated by modifying the mode-coupling expression for the 
viscosity near equilibrium\cite{banchio1999b}. 
This can be done by following the same procedure explained in Section
\ref{sec:mct} for the total momentum both from solvent molecules and
colloids.  
The equation for the total momentum is given by the Navier-Stokes
equation. 
The nonlinear term in the equation comes from the osmotic pressure of 
the suspension particles. 
Neglecting the coupling of the density field of colloids with that of 
the solvent molecules, one obtain the mode-coupling expression for the
viscosity. 
In the presence of the shear, the same modification as in 
eq.(\ref{eq:mct.deltazeta3}) is necessary and the wave vectors should be
replaced by their time-dependent counterparts. 
The final result is given by 
\begin{equation}
\begin{aligned}
&
\eta(\dot{\gamma})
=
\eta_0
+
\frac{1}{2\beta}
\\
&
\times
\int_{0}^{\infty}\!\!\mbox{d} t
\int\!\!\frac{{\mbox{d}}{\bf k}}{(2\pi)^2}
\frac{k_{x}k_{x}(t)}{S^2(k)S^2(k(t))}
\frac{\partial S(k)}{\partial k_{y}}
\frac{\partial S(k(t))}{\partial k_{y}(t)}
F^2({\bf k}(t),t)
,
\end{aligned}
\end{equation}
where $\eta_{0}$ is the viscosity of the solvent alone. 
The integral over ${\bf k}$ 
can be implemented for the set of $F(k,t)$ evaluated using 
eq.(\ref{eq:mct.mctfinal}). 
In Fig. 14, 
we have plotted the shear dependence of the reduced viscosity defined by 
\begin{equation}
\eta_{R}(\dot{\gamma})\equiv 
\frac{  \eta(\dot{\gamma}) -\eta_{0} }{\eta_{0}}
\label{eq:results.eta}
\end{equation}
for various densities around $\phi_c$.
\begin{figure}[ht]
\includegraphics[scale=0.6,angle=0]{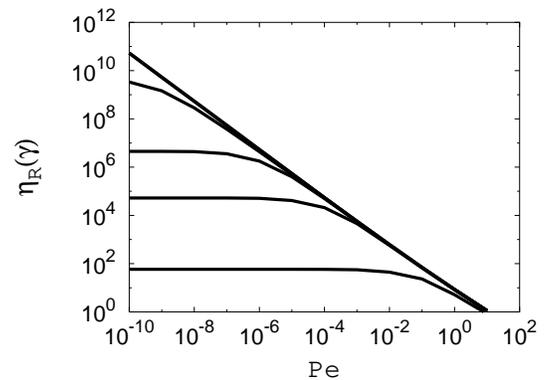} 
\caption{The reduced viscosity, eq.(\ref{eq:results.eta}), 
versus the shear rate for various densities.   
Here $\mbox{Pe} =\dot{\gamma}\sigma^2/D_{0}$. 
From top to bottom;
$\phi = 0.766549$, $0.76650$, $0.766453$, $0.76640$, $0.76600$, and $0.75600$.
The highest density is $4\times 10^{-5}$ \% larger than $\phi_c$. 
} 
\label{fig:visc}
\end{figure}
The strong non-Newtonian behavior is observed at high shear rate and
large densities, which is again in qualitative agreement with the
simulation results for liquids reported in Ref.\cite{yamamoto1998c}. 
The shear thinning exponent extracted from the data between 
$10^{-10}~ < ~\mbox{Pe} ~ < 1$
is 
$\eta_{R}(\dot{\gamma})\propto \dot{\gamma}^{-\nu}$ with $\nu \simeq 
0.99$. 
This is in agreement with the exponent estimated from the simulation results
and 
the structural relaxation time, $\tau_{\alpha}(\dot{\gamma})$, observed
in the simulation and in the theory discussed in the previous
subsections. 
For larger shear rates, Pe $>1$, the exponent becomes smaller,
which is again consistent with computer 
simulation\cite{yamamoto1998c}. 
However, one should not trust this reasoning for Pe $>$1. 
As discussed in Subsection \ref{subsec:Sk}, 
it is expected that the distortion 
of structure $c(k)$ and $S(k)$ by shear becomes important and one should
take these effects into account in the theory. 
In this regime, however, 
the glassy structure has already been destroyed
and the system become more like a ``liquid''.
In the liquid state, 
the shear-thinning exponent is always expected to be smaller
than 1 and range between 0.5 and
0.8\cite{naitoh1976,kirkpatrick1985,barrat2000,vanderwerff1989c}. 

Slightly above $\phi_{c}$, plastic behavior is observed, 
which implies the presence
of the yield stress.
Note that this is an artifact of the theory because 
$\phi_{c}$ predicted from the mode-coupling theory is  much lower than the
real glass transition density. 
Below the real glass transition density $\phi_g$, 
it is expected that shear thinning behavior similar to that predicted by
MCT for $\phi < \phi_c$ will result.

\section{Conclusions}
\renewcommand{\theequation}{\Roman{section}.\arabic{equation}}  
\setcounter{equation}{0}

In this paper, we presented 
the derivation of the mode-coupling equation for the 
{\it realistic}
supercooled fluids 
under shear and compared the results with the molecular dynamic
simulation. 
Our starting point is fluctuating hydrodynamics extended to the 
molecular length scales. 
A simple closed equation  for the intermediate
scattering function was derived. 
We applied the theory to a two-dimensional colloidal suspension with
hard-core interactions. 
The numerical analysis of the equation revealed very good agreement 
with simulation results for a related system: a binary liquid
interacting with a soft-core potential. 
Theory and simulation showed common features such as 
(i) drastic reduction of relaxation times and the viscosity 
and 
(ii) nearly isotropic relaxation irrespective of the
direction of the shear flow. 
The fact that the dynamics is almost isotropic supports
the validity of the schematic models proposed so
far, in which the anisotropic nature of the
nonequilibrium states was not explicitly
considered\cite{berthier2000b,fuchs2002}.   

The mode-coupling theory developed in this paper is far from complete. 
The most crucial approximation is the use of the 1st FDT, which was
employed when we close the equation for dynamical correlators. 
It is already known that the 1st FDT is violated for supercooled 
systems under shear as well as aging systems\cite{barrat2000}. 
Without the 1st FDT, one has to solve simultaneously
the set of mode-coupling equations
for the propagator and correlation function, which couple each other
through the memory kernels. 
Research in this direction is under way.
A simple argument, however, should suffice to explain the success of the
present theory with regard to shear thinning behavior. 
Consider a system that evolves out of equilibrium with one effective
temperature $T_{\mbox{\scriptsize eff}}$. 
Crudely, we can use the effective steady state version of FDT violation
in the form
\begin{equation}
\chi_{ij}(t)
=-\frac{\theta(t)}{k_{\mbox{\scriptsize B}} T_{\mbox{\scriptsize eff}}}
\frac{\mbox{d} C_{ij}(t)}{\mbox{d} t } 
\label{eq:conc.1stFDT}
\end{equation}
to eliminate the response function $\chi_{ij}(t)$ in favor of the
correlation function $C_{ij}(t)$ from the set of mode-coupling
equations. 
We thus expect for the behavior of the structural relaxation time,
$\tau_{\alpha}(\dot{\gamma})$, will be unaffected as long as
eq.(\ref{eq:conc.1stFDT}) holds. 
Another important approximation was to neglect 
the small distortion of the structure ($c(k)$ and $S(k)$) due to
shear. 
The simulation supports that the distortion is negligibly small at
the low shear rates (small P\'{e}clet number). 
The construction of the equation for equal-time correlation
functions such as the structure factor 
might be more subtle and should be considered in future.
Simulation results show that the structure is distorted in a noticeable
and anisotropic manner at 
high shear rates;
the peak height of the first peak of $S(k)$ was lowered as much as 10 percent
at $\dot{\gamma} = 10^{-1}$. 
It is interesting that the dynamics is still isotropic and
qualitative behavior is not affected even for such high shear rates. 

In the absence of shear, 
MCT is known to break down at densities 
well below the real glass-transition density, where
MCT incorrectly predicts a fictitious non-ergodic transition. 
Beyond this MCT cross-over density, activated hopping
between the local minima of the free energy surface is expected to
dominate the dynamics of the system. 
Sollich {\it et al.}\cite{sollich1997,fielding2003} have analyzed 
a schematic model for hopping processes and explained
similar shear thinning behavior. 
Lacks\cite{lacks2001} has also 
rationalized shear thinning behavior in terms of changes
of the free energy barriers due to shear.
It is interesting that the totally different picture given by MCT 
leads to some qualitatively similar conclusions. 
The analysis of the violation of the 1st FDT and effective temperatures
has the possibility of clarifying the difference between these distinct
pictures.   
Barrat {\it et al.} have shown by simulation that the 1st FDT is violated
in a sheared liquid and observed non-trivial effective
temperatures\cite{barrat2000}.  
This is consistent with the conclusion of 
Berthier {\it et al.}\cite{berthier2000b} for the non-equilibrium $p$-spin
model. 
On the other hand, the trap model predicts multiple effective
temperatures\cite{fielding2003}.  
Future effort will be directed towards extracting effective temperatures
from our fluctuating hydrodynamic approach.

In Section II we have mentioned 
that there are subtle problems in constructing an equation
for the intermediate scattering function alone:
The ``correct'' MCT equation can be derived  if you start from the 
nonlinear Langevin equations for the density and momentum fields,  
taking the overdamped limit at the end. 
But difficulties arises if the overdamped limit is taken for the Langevin
equation at the beginning. 
These problems exist even for systems at equilibrium and 
are related to generic problems that exist in loop
expansion methods. 
These subtleties were already recognized in the mid-70's\cite{deker1975}
and recently pointed out in the context of glassy systems 
by Schmitz {\it et al.}\cite{schmitz1993}. 
A detailed study of these and 
issues related to the field-theoretic approach to out-of-equilibrium
glassy liquids will be discussed in a future publication\cite{miyazaki2003b}.

\begin{acknowledgments}
The authors acknowledge support from NSF grant \#0134969.  
The authors would like to express their gratitude to 
Prof. Matthias Fuchs for
for useful discussions. 
KM and RY would like to thank A. Onuki for helpful discussions.
\end{acknowledgments}

\appendix
\section{Static structure for the two-dimensional fluids with
the hard-core interaction} 
\renewcommand{\theequation}{\Alph{section}.\arabic{equation}}  
\setcounter{equation}{0}

Baus {\it et al.} have derived an
approximated analytic expression of the direct
correlation function $c(k)$ for the $d$-dimensional hard sphere 
fluids\cite{baus1987}. 
For the two dimension system,  it is given by
\begin{equation}
\begin{aligned}
&
nc(k) 
= 
-\phi
\frac{\partial \{\phi Z(\phi)\}}{\partial \phi } 
\left[
4\left( 1-a^2\phi\right)f(k\sigma)
\right.
\\
&
\left.
+a^{2}\phi 
\left\{ a^{2}\left( \frac{J_1(ak\sigma/2)}{ak\sigma/2}\right)^2
\right.
\right.
\\
&
\left.
\left.
+ 
\frac{16}{\pi}
\int_{1/a}^{1}\!\!{\mbox{d}} x~
(1-x^2)^{1/2}
\left(
\frac{J_1(k\sigma)}{k\sigma}-(ax)^{2}\frac{J_1(ak\sigma x)}{ak\sigma}
\right)
\right\}
\right]
,
\end{aligned}
\end{equation}
where $J_n(x)$ is the Bessel function of the first kind, 
$\phi = \pi \sigma^2 \rho_{0}/4$ 
is the packing fraction, $\rho_{0}$ is 
the number density, and $\sigma$ is the diameter of spheres.
$a$ is a parameter which is determined by solving
\begin{equation}
\begin{aligned}
&
\frac{2}{\pi}
\left[
a^2(a^2-4)\arcsin\left(\frac{1}{a}\right)
 -\left(a^2+2\right)\sqrt{a^2-1}
\right]
\\
&
= \frac{1}{\phi^2}
\left[
1-4\phi-\left\{ 
\frac{\partial \{\phi Z(\phi)\}}{\partial \phi } 
 \right\}^{-1}
\right]
.
\end{aligned}
\end{equation}
$Z(\phi)\equiv p/\rho_{0} k_{\mbox{\scriptsize B}} T$ is the compressibility factor
which is expanded by {\it a rescaled virial series} as
\begin{equation}
Z(\phi) = (1-\phi)^{-2}\left( 1 + \sum_{n=1}^{\infty}c_{n}\phi^{n}\right),
\end{equation}
where $c_n$ is a coefficient which is related to the virial coefficients
via a recurrence relation. 
We truncated the series at $n=6$. 
The coefficients $c_{n}$ are given by 
$c_1=0$, $c_2=0.1280$, $c_3=0.0018$, $c_4=-0.0507$, $c_5=-0.0533$, and 
$c_6=-0.0410$.

\end{document}